\documentclass[11pt,twoside]{article}
\usepackage{aspen2004pulsars}
\usepackage{psfig}
\begin{document}
\title{Pulsars in Globular Clusters}
\author{Fernando Camilo}
\affil{Columbia Astrophysics Lab, Columbia University, 550 West 120th Street,
New York, NY 10027, USA}
\author{Frederic A.\ Rasio}
\affil{Department of Physics and Astronomy, Northwestern University,
2131 Tech Drive, Evanston, IL 60208, USA}

\begin{abstract}
More than 100 radio pulsars have been detected in 24 globular clusters.
The largest observed samples are in Terzan~5 and 47~Tucanae, which
together contain 45 pulsars.  Accurate timing solutions, including
positions in the cluster, are known for many of these pulsars.  Here we
provide an observational overview of some properties of pulsars in
globular clusters, as well as properties of the globular clusters
with detected pulsars.  The many recent detections also provide a new
opportunity to re-examine theoretically the formation and evolution of
recycled pulsars in globular clusters. Our brief review considers the
most important dynamical interaction and binary evolution processes:
collisions, exchange interactions, mass transfer, and common-envelope
phases.
\end{abstract}

\section{Introduction}

Following the discovery of the first millisecond pulsar (MSP) by Backer et
al.\ (1982) and the wide acceptance of the ``recycling'' model, connecting
low-mass X-ray binaries (LMXBs) to MSPs (Alpar et al.\ 1982), globular
clusters (GCs) became a favorite place to search for MSPs. Indeed, GCs
were known to contain surprisingly large numbers of LMXBs (Clark 1975),
which should produce MSPs as their accreting neutron stars (NSs) are spun
up and recycled into fast radio pulsars.  After a flurry of activity,
the first pulsar in a globular cluster (the single PSR~B1821$-$24,
with period $P=3$\,ms) was found in M28 by Lyne et al.\ (1987).

In the following 17 years much has been learned observationally and
theoretically about the population of pulsars in GCs, and we provide
here a brief review taking into account the latest detections. Note
especially that, after the Aspen conference was held in January 2004,
but before this review was completed (several months later), many more
GC pulsars were discovered, including the 20 new MSPs detected in Ter~5
by Ransom et al.\ (2005).  We have included these recent discoveries in
our review (see Table~1) in order to provide an up-to-date summary.

For other reviews, somewhat dated but still excellent, see Kulkarni \&
Anderson (1996) and Phinney (1996), both written at a time when the
total number of GC pulsars was only about 30!

\begin{figure}[t]
\centerline{
\psfig{file=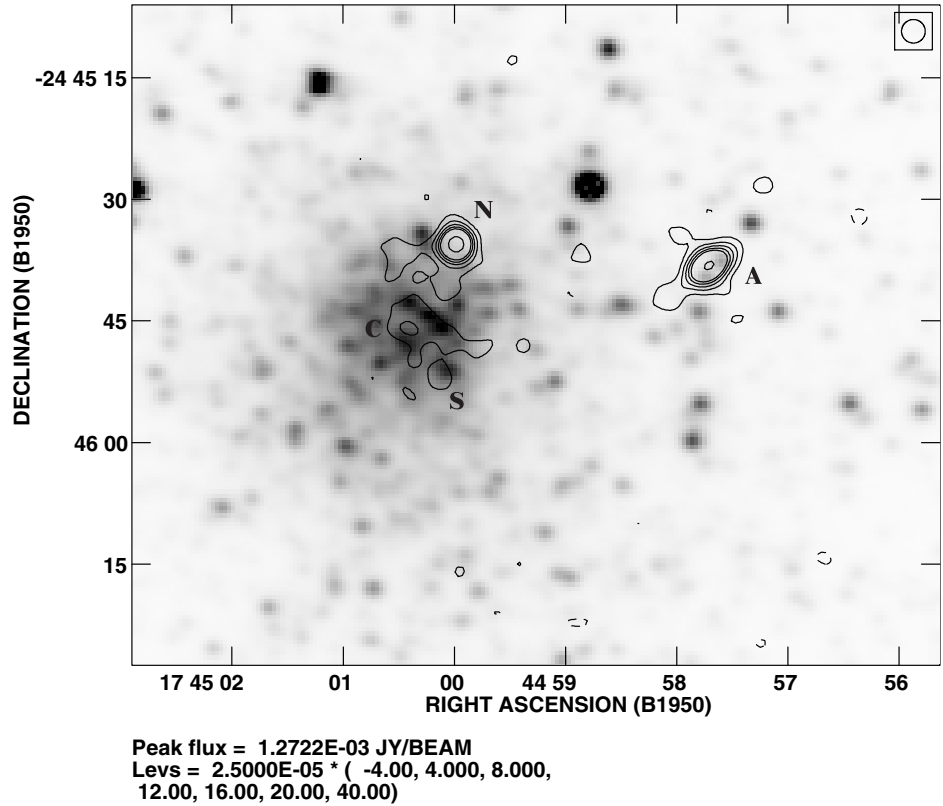,width=0.93\textwidth}}
\caption{Terzan~5 imaged with the VLA at 20\,cm wavelength (Fruchter
\& Goss 2000), overlaid on an optical $I$-band image.  ``A'' marks the
strong radio emission from PSR~J1748--2446A (Ter~5~A), the first pulsar
discovered in this globular cluster.  Near the cluster core there is
substantial diffuse emission with a steep spectrum, possibly originating
from unidentified pulsars.  The 20 new pulsars recently detected in
Ter~5 by Ransom et al.\ (2005) presumably account for a significant
portion of this emission.  }
\end{figure}

\section{Searches for Pulsars in Globular Clusters}

\subsection{Previous Searches}

A powerful technique used to determine good GC targets for pulsation
searches relies on imaging GCs to find those which contain steep
spectrum, possibly polarized, radio sources --- i.e., having the
properties of pulsars (see Fig.~1).  Alternatively, one can proceed
straight to searching for pulsations from ``all'' globular clusters,
starting with those that are closest to the Earth or have the smallest
predicted dispersion measures (DMs).  Once a first pulsar is discovered,
the approximate DM for other pulsars in the same GC is known, simplifying
considerably the search process. In this manner, about 30~pulsars had
been found in 12~GCs by 1992.

\begin{figure}[t]
\centerline{
\psfig{file=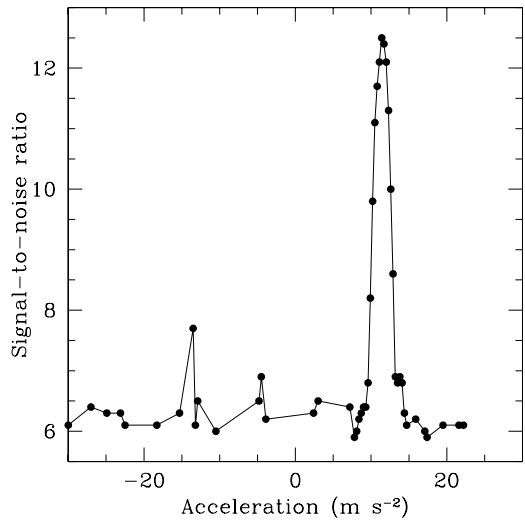,width=0.54\textwidth}}
\caption{Signal-to-noise ratio vs.\ trial acceleration for the discovery
of 47~Tuc~R, a binary MSP with the shortest orbital period known for
a pulsar, $P_b = 96$\,min.  This pulsar was discovered in 17\,min
sub-integrations at a frequency of 1.4\,GHz with the Parkes telescope
and would not have been detected without computationally onerous
``acceleration searches'' (Camilo et al.\ 2000).  It has also only ever
been detected on one day, showing the effect of interstellar scintillation on
detectability for some pulsars. }
\end{figure}

Searches resumed in earnest in 1998, using a combination of upgraded/new
telescopes, frequencies, spectrometers, and analysis techniques
(particularly, widespread  ``acceleration searches''; see Fig.~2), and,
by the end of 2004, about 100~pulsars were known in 24~GCs.

\subsection{On-going Searches}

{\bf Arecibo:} Twenty-two GCs within 50\,kpc are being searched at
1.4\,GHz (see Hessels et al.\ 2004 and Ransom et al., in this volume).

\medskip

\noindent {\bf GBT:} Seven GCs were searched at 1.4\,GHz by Jacoby et al.\
(2002), and about 13 more are being done at 1.4 and 2.0\,GHz (Ransom et
al.\ 2004, 2005; also in this volume; Hessels et al.\ 2004).

\medskip

\noindent {\bf GMRT:} About 10 GCs are being searched with $\sim 2$\,hr
integrations at 0.3\,GHz (see Freire et al.\ 2004; also in this volume).

\medskip

\noindent {\bf Parkes:} In addition to 47~Tuc, with 22 pulsars known,
60 other GCs with predicted $\mbox{DM} \la 300$\,cm$^{-3}$\,pc are being
searched with $\sim 2$\,hr integrations at a frequency of 1.4\,GHz,
with work on 45 of these essentially complete (see Possenti et al.,
in this volume).

\medskip

Collectively, these searches are proving to be very successful.  However,
as we survey here the properties of the pulsars, and the clusters they
reside in, it is important to keep in mind that several selection effects
are present:

\begin{enumerate}

\item Distance: we only detect the most luminous pulsars from many GCs.

\item DM/$P$ and acceleration: it is more difficult to detect pulsars
with larger DM and/or shorter periods, and particularly difficult to
detect MSPs in very tight binaries.

\item Data processing: very significant amounts of computing (and people)
power are required to analyze the data sets completely --- e.g., years
after being collected, 47~Tuc data are now being analyzed with greater
sensitivity than previously done to pulsars having smaller $P$ and $P_b$.

\item Propagation effects in the ISM: for some clusters it is necessary to
observe often in order to detect weak pulsars on scintillation maxima.
The same may be true for some pulsars that display eclipses.  For some
GCs, multi-path propagation (scattering) may prevent detection of
short-period pulsars at relatively low observing frequencies ($\la
1.5$\,GHz).

\item Pulsar spectra: ideally, multiple observing frequencies should
be used.

\end{enumerate}

\section{Properties of Pulsars in Globular Clusters}

One hundred pulsars with reasonably well-measured parameters are
known in 24~GCs as of this writing.  Some important parameters of
both the GCs and the pulsars are listed in Table~1. An updated table
of GC pulsar parameters is maintained online by P.\ Freire at {\tt
http://www.naic.edu/$\sim$pfreire/GCpsr.html}.

\subsection{Spin Period Distribution}

\begin{figure}[t]
\centerline{
\psfig{file=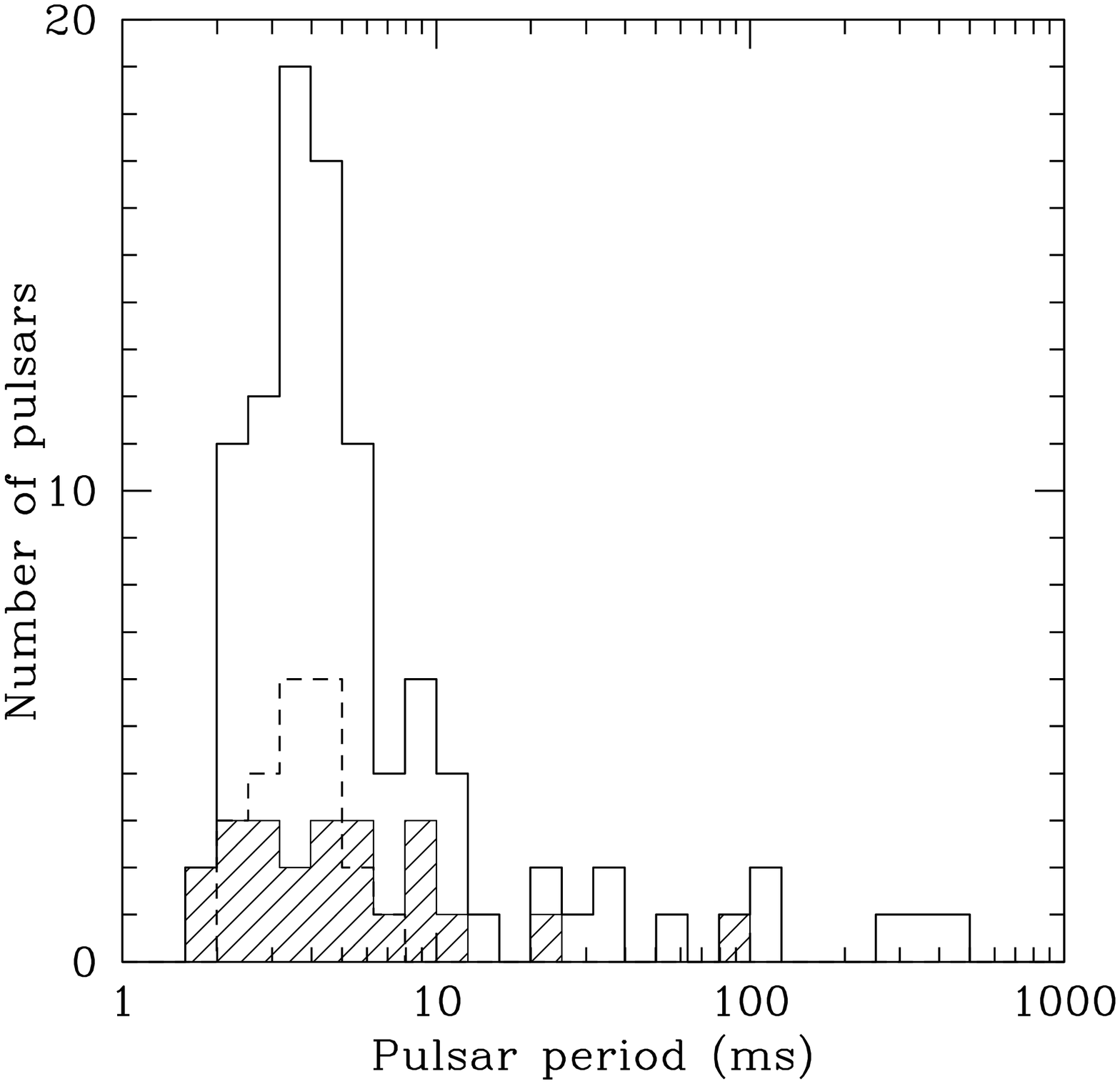,width=0.50\textwidth}
\psfig{file=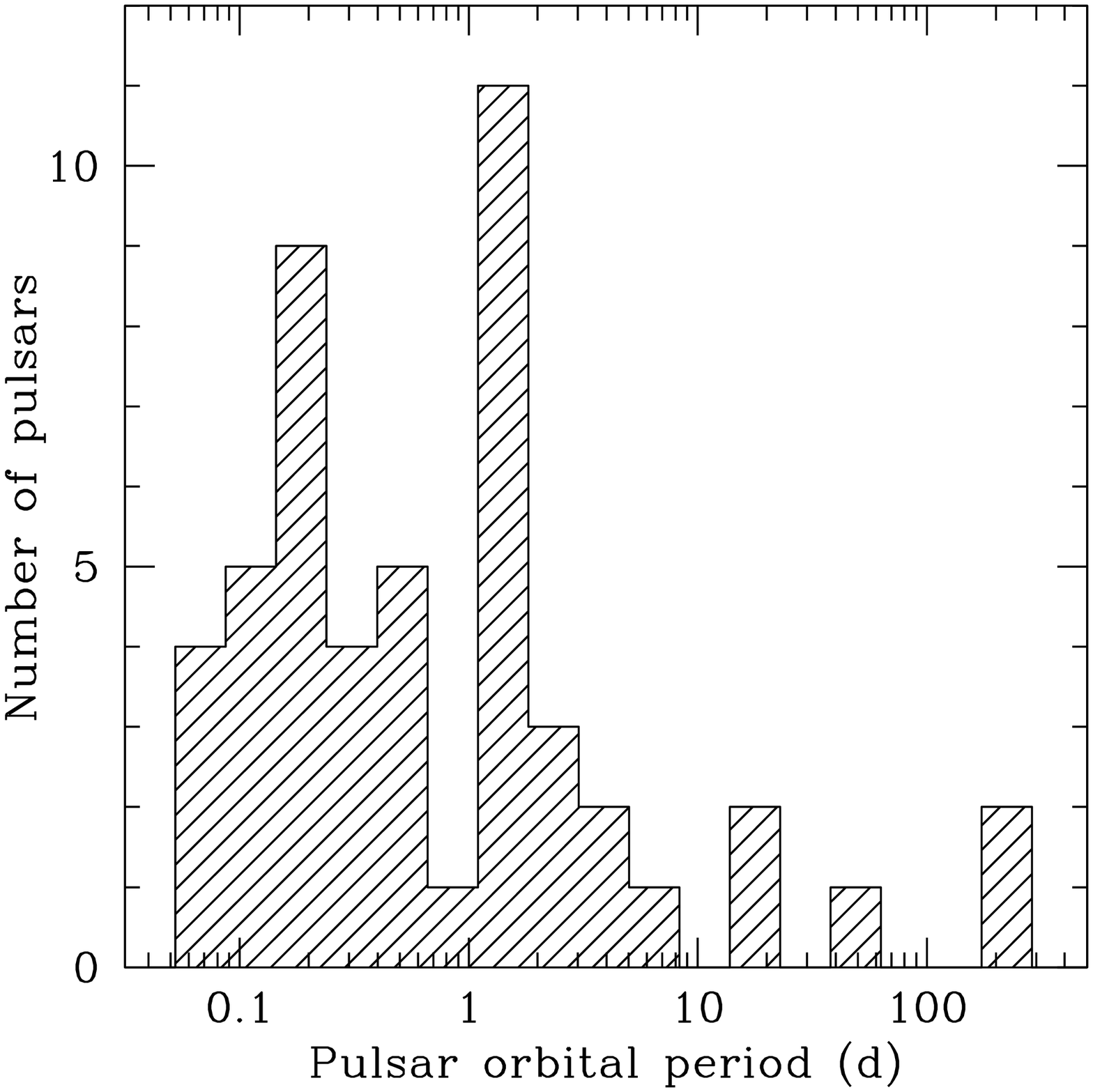,width=0.50\textwidth}}
\caption{\textit{Left:} Distribution of spin periods for pulsars known
in globular clusters.  The shaded area represents the 23~pulsars in
Ter~5, while the 22 pulsars known in 47~Tuc are shown by the dashed-line
histogram. \textit{Right:} The distribution of orbital periods for 50
binary pulsars in globular clusters. }
\end{figure}

\noindent As shown in Fig.~3 (left), the vast majority of pulsars in GCs
are MSPs.  There are only three slow (``young'') pulsars in this set,
although these may have a high birth rate (Lyne, Manchester \& D'Amico
1996).  As shown by the dashed-line histogram in the figure, the 47~Tuc
pulsars have a very narrow distribution, and none is known at $P<2$\,ms,
despite considerable sensitivity down to $P\la 1$\,ms.  By contrast, the
distribution of the pulsars known in Ter~5 (shaded histogram) is broader.

\subsection{Binary Period Distribution}

The binary period distribution for pulsars in GCs is shown in Fig.~3
(right).  Fifty of the 100 pulsars known in GCs are represented here.
Nine more are in binaries with periods yet to be determined, and it
must be remembered that short-period binaries are selected against
in pulsation searches.  Still, it appears that the fraction of single
pulsars in GCs ($\sim40\%$ observed) is larger than the corresponding
fraction of single MSPs in the Galactic disk ($\sim20\%$).

There are evidently two main populations of pulsars distinguished by
orbital period: those with periods of a few hours, and those with $P_b
\sim 1$--2\,d with a tail extending to $\ga 10$\,d.  Two pulsars with $P_b
> 100$\,d stand out from this trend: PSR~B1620$-$26, a triple system in M4
(e.g., Thorsett et al.\ 1999; Sigurdsson \& Thorsett, in this volume),
and PSR~B1310+18 in M53 (see \S 6.3).

\subsection{Pulsar Companions}

The two groups selected by binary period correspond roughly to those
pulsars, often eclipsing, that have $m_2 \sim 0.03\,M_\odot$ dwarf
companions (the short-$P_b$ systems), and those with $\sim 0.2$\,$M_\odot$
``He white dwarf'' (WD) companions (see Fig.~4, left).  In most cases
these statements are drawn by analogy with the kinds of binary pulsar
systems known in the Galactic disk, based on the measured mass function:
in few cases do we actually detect the pulsar companions directly, for
instance via optical emission.  Also, by comparison with the situation
in the disk, there is an apparent dearth of massive (CO) WDs among
the GC pulsars; the recently discovered Ter~5~N (Ransom et al.\ 2005;
Table~1) could be one such example.  In addition to these groups, a very
few pulsars have NS companions.  Finally, importantly, some GC pulsars
have companions with no analogue in the disk: systems that show eclipses
and have $m_2 \ga 0.1\,M_\odot$, which we designate here generically as
``main sequence'' (MS) companions.

In the case of 47~Tuc, the known pulsars are divided roughly into 1/3
isolated, 1/3 with $\sim 0.03$\,$M_\odot$ companions, 1/3 with $\sim
0.2$\,$M_\odot$ companions, and one with a $\sim 0.1$\,$M_\odot$ MS
companion, 47~Tuc~W (Edmonds et al.\ 2002).

\begin{figure}[t]
\centerline{
\psfig{file=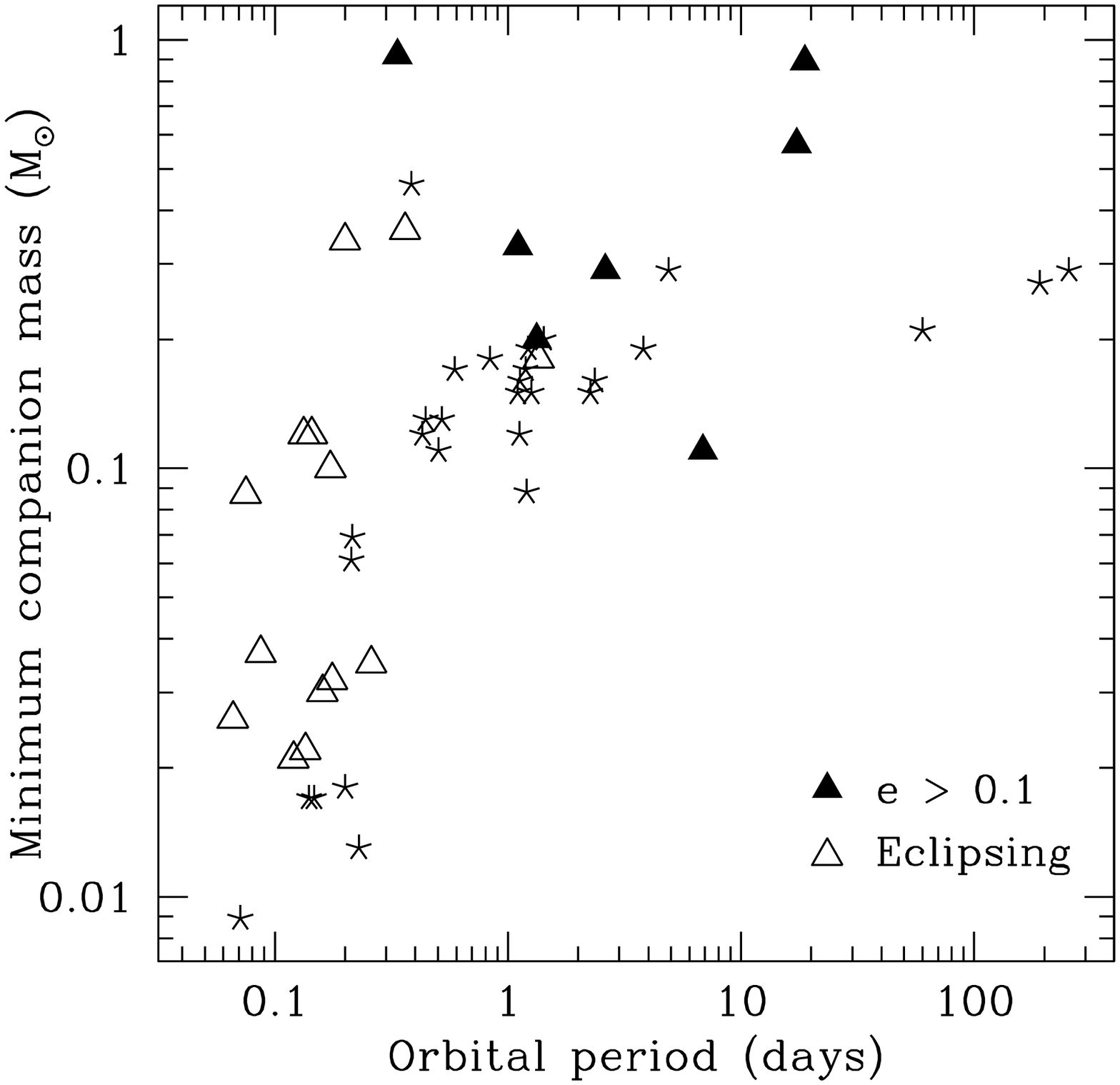,width=0.50\textwidth}
\psfig{file=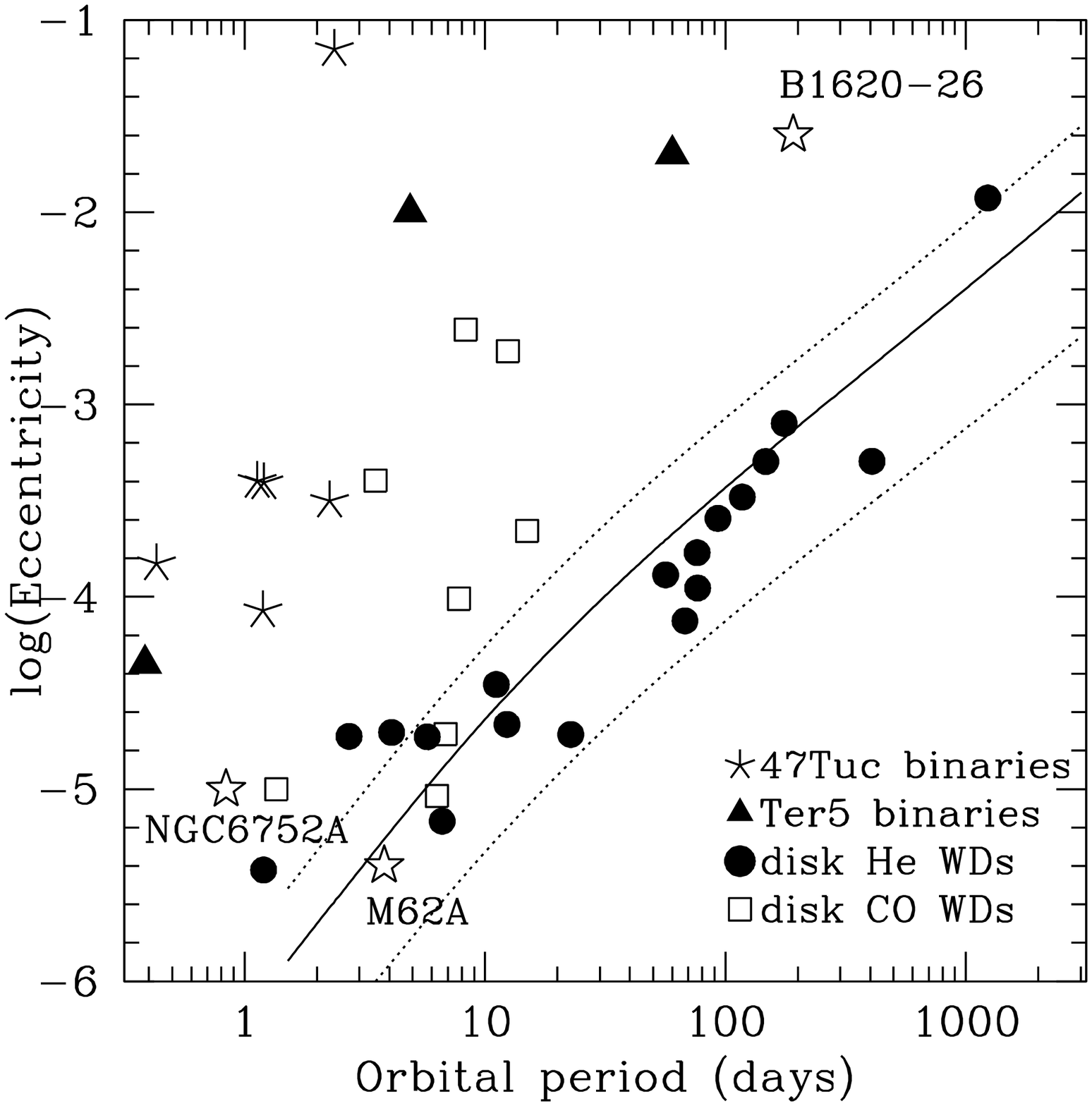,width=0.50\textwidth}}
\caption{\textit{Left:} Minimum companion mass (derived from the mass
function assuming a pulsar mass of 1.35\,$M_\odot$ and orbital inclination
of $i=90\deg$) vs.\ binary period for pulsars known in globular clusters.
Pulsars that are known to eclipse or have eccentricities $e >0.1$ are
indicated.  \textit{Right:} Eccentricity vs.\ orbital period. Field
pulsars with low-mass companions that are thought to have formed via
stable mass transfer (denoted as having ``He WD'' companions) follow
the prediction of the fluctuation--dissipation model of Phinney (1992),
represented by the solid line and 95\% confidence level dashed lines.
Galactic disk pulsars with more massive CO WD companions do not follow
this trend.  Most GC pulsars with low-mass companions here (those in
47~Tuc, Ter~5 and M4) also do {\em not\/} follow those predictions:
their eccentricities are unusually ``large'' (even when $e \sim 10^{-4}$).
We note that the Ter~5 pulsars indicated here have more massive companions
than those in 47~Tuc (see Table~1).  It is also notable that two of
the GC pulsars (in M62 and NGC~6752) {\em do\/} have extremely small
eccentricities.  }
\end{figure}

\subsubsection{Eclipses}

Fig.~4 (left) shows that most of the 14 eclipsing systems known in GCs
have $2 \la P_b \la 6$\,hr. Half of these have $\sim 0.03\,M_\odot$
companions like the ``black widow'' eclipsing binaries PSRs~B1957+20
and J2051$-$0827 of the Galactic disk.  Interestingly, five otherwise
similar systems with $0.01 \la m_{2\,\rm min} \la 0.02\,M_\odot$ do
{\em not\/} show eclipses, suggesting that they have companion masses
similar to those of the eclipsing binaries, but are viewed in a more
face-on geometry.  Six other eclipsing binaries are unusual in that their
companions are substantially more massive ($m_2 \ga 0.1$\,$M_\odot$):
Ter~5~A, 47~Tuc~W, M62~B, M30~A, and in particular Ter~5~P and 47~Tuc~V
with $m_2 \ga 0.3$\,$M_\odot$.  One system is further unusual in having a
larger orbital period: PSR~J1740--5340 in NGC~6397, with $P_b =1.3$\,d,
variable eclipses (D'Amico et al.\ 2001), a $\simeq 0.25$\,$M_\odot$
(Ferraro et al.\ 2003) ``red straggler'' companion showing ellipsoidal
optical variations (Orosz \& van Kerkwijk 2003), and possibly X-ray
variability (Grindlay et al.\ 2002).  See the article by Freire, in this
volume, for more on eclipsing pulsars.

\subsubsection{Eccentricities}

Fig.~4 (right) shows the orbital eccentricity of binary pulsars vs.\ their
orbital period.  It is clear by comparison with otherwise equivalent
pulsars in the Galactic disk that even the ``small'' eccentricities
of most GC binaries are unusually large --- a clear sign of stellar
interactions either during or post formation (\S 6).  Surprisingly, the
eccentricity of the binary M62~A, located near the core of a very dense
GC, is extremely small (as is that of NGC~6752~A, which, in contrast,
is located well outside the core of its parent GC).

Also in Fig.~4 (left) we indicate seven of the eight known GC pulsar
systems having very large eccentricities, $e >0.1$.  These are: M15~C,
a NS--NS system likely formed in a 3-body exchange and ejected out of the
GC core (Prince et al.\ 1991; Phinney \& Sigurdsson 1991); B1802$-$07
in NGC~6539 (Thorsett et al.\ 1993), with perhaps a He WD companion
(formed through a collision with a red giant; see \S 6.2), or even a
MS companion; B1516+02B in M5, about which not much is known (not even
whether its companion is a WD or a NS; Anderson et al.\ 1997; see \S
6.3); J0514$-$4002 in NGC~1851, a 5\,ms pulsar in a 19\,d orbit of $e =
0.9$, with a massive companion whose nature is unclear (Freire et al.\
2004; also in this volume); J1750$-$37 in NGC~6441, with a relatively
large $P=111$\,ms and $m_2 \ga 0.5$\,$M_\odot$ (Possenti et al.\ 2001;
also in this volume); and the recently discovered Ter~5~I ($P=9$\,ms)
and Ter~5~J ($P=80$\,ms), each with $P_b \simeq 1$\,d, $e \simeq 0.4$,
and a total system mass of $\simeq 2.2\,M_\odot$, derived from the
measured advance of periastron (Ransom et al.\ 2005; see also \S 6.2).
In addition, M30~B has $e \ga 0.5$ (Table~1).

\subsection{Radial Distribution in Clusters}

Fig.~5 shows the distribution of pulsar--GC center angular offset,
in terms of the GC core radius $r_c$.  The vast majority of pulsars
are located at $0.2 < r/r_c < 5$.  It should be noted that this is a
snapshot in time and that some pulsars are located in eccentric orbits
about their cluster centers (\S 6.3), although the bias of this effect
for the population as a whole is likely small.

\begin{figure}[h]
\centerline{
\psfig{file=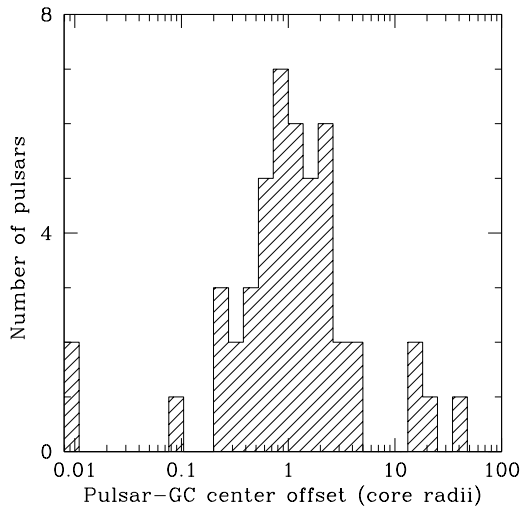,width=0.52\textwidth} }
\caption{Histogram for 48 pulsar--globular cluster center angular offsets,
in units of core radius $r_c$. }
\end{figure}

The four exceptions to this high degree of central concentration for the
pulsars are: the ``unusual'' PSR~J1740$-$5340 in NGC~6397; the NS--NS
binary M15~C; and two otherwise normal MSPs (one isolated) in NGC~6752
(D'Amico et al.\ 2002).

We now comment on the case of PSR~B1718$-$19, a very unusual eclipsing
pulsar with $P = 1$\,s, sometimes assumed to be associated with NGC~6342
(Lyne et al.\ 1993).  The main argument against an association is that
for this pulsar $r = 46\,r_c$.  But, as can be seen from Fig.~5, this
would no longer be a uniquely large offset.  It is also curious that the
metallicity of NGC~6342 is high, much like those of other GCs where slow
pulsars are located.  See also Bailes et al., in this volume.

\section{Cluster Properties}

\subsection{What Globular Clusters Have Pulsars?}

Fig.~6 (left) shows a scatter plot of metallicity vs.\ central density for
70 GCs (out of about 150 known in the Galaxy) that have been searched at
{\em some\/} level (many of the searches are not yet complete, and the
luminosity limits vary widely).  It can be seen from the figure that
pulsars are known in GCs with metallicities in a wide range and with
$\rho_0 > 10^3$\,$L_\odot$\,pc$^{-3}$.

\begin{figure}[h]
\centerline{
\psfig{file=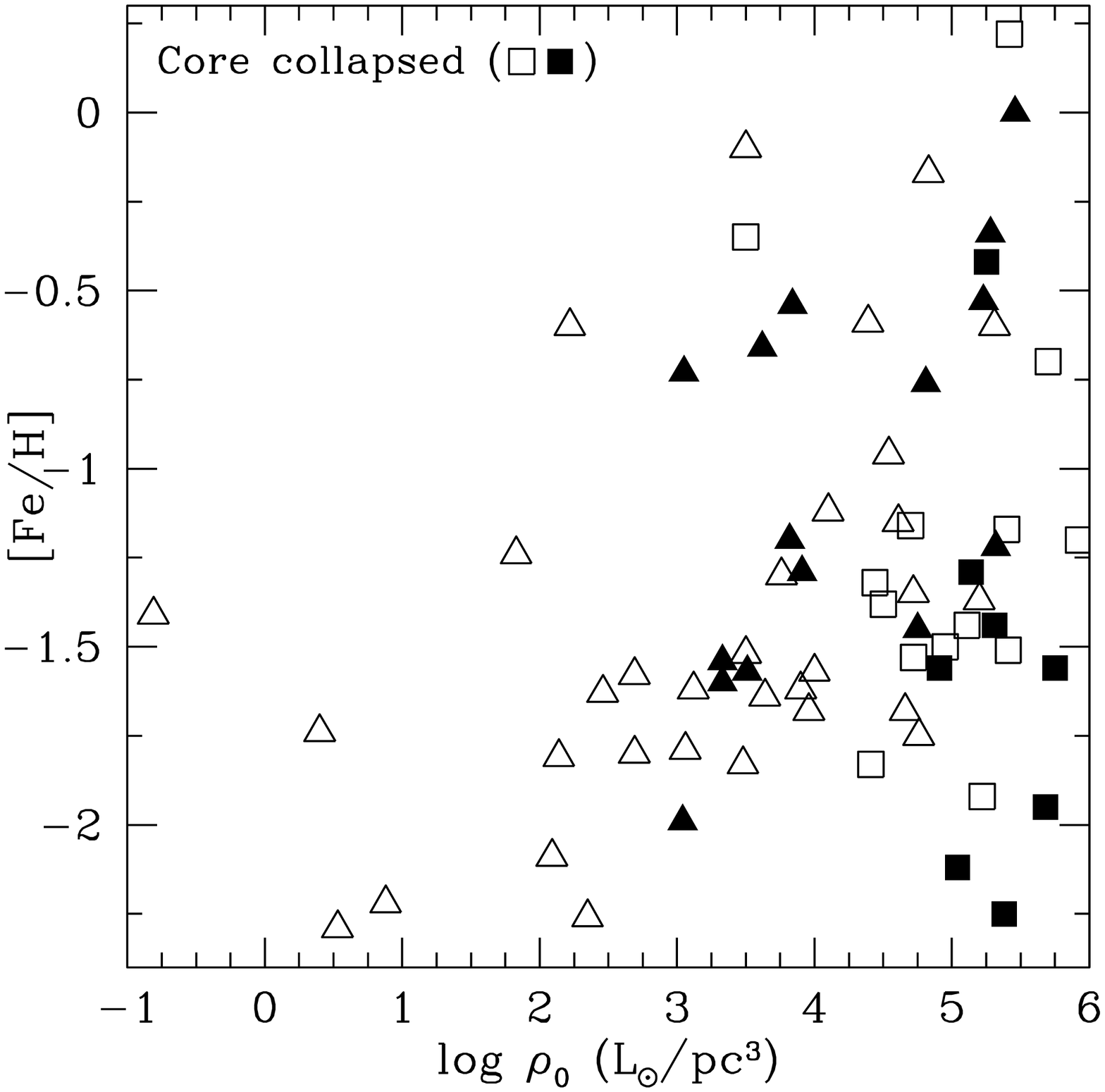,width=0.50\textwidth}
\psfig{file=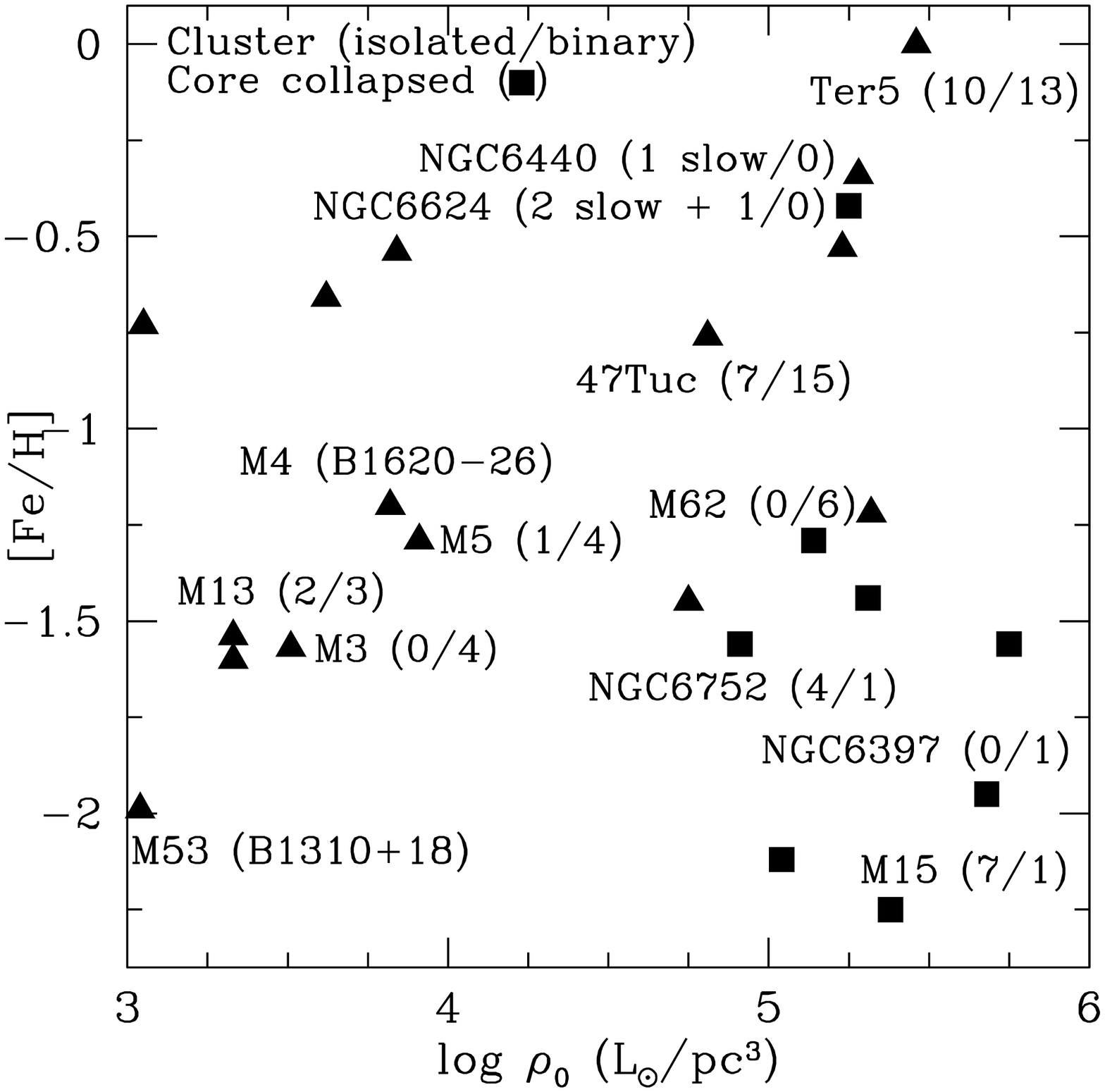,width=0.50\textwidth}}
\caption{\textit{Left:} Metallicity vs.\ central density for GCs
searched at some level for pulsars.  Core-collapsed clusters are
indicated (squares), as are those where pulsars have been found (filled
symbols). \textit{Right:} A zoom-in on the left panel showing only those
clusters having known pulsars.  Some particular clusters are also named,
with an indication of the number of isolated and binary pulsars known
in each. }
\end{figure}

A word on sensitivity puts Fig.~6 (left) into perspective: at Parkes,
in all seven GCs with recent pulsar discoveries, the luminosity at
1400\,MHz of the {\em brightest\/} pulsar in the GC is within a factor
of two of $L_{1400} \approx 12$\,mJy\,kpc$^2$.  On the other hand,
quite often the luminosity {\em limit\/} for various Parkes searches is
$\ga 10$\,mJy\,kpc$^2$.  Clearly, non-detection of a pulsar at these
luminosity levels need not engender even the thought that pulsars do
not exist in such GCs!  Searches using the GBT or Arecibo can be more
sensitive, as exemplified by the recent discovery with the GBT at a
frequency of 2.0\,GHz of 20 pulsars in Ter~5 (Ransom et al.\ 2005;
see also Table~1), a cluster previously extensively searched at Parkes.

\subsection{Clusters with Pulsars}

Fig.~6 (right) shows a scatter plot of the metallicity vs.\ central
density for GCs in which pulsars are known.  Although the statistics
are poor (and nearly half of all pulsars known reside in either Ter~5
or 47~Tuc), some tentative statements can be made:

\begin{itemize}

\item Long orbital period pulsars reside predominantly in low-density GCs.

\item Pulsars with large $r/r_c$ are located in high-density GCs.

\item Pulsars with ``MS'' companions are located in high-density GCs.

\item Slow pulsars are located in high-metallicity, high-density GCs.

\item The binary fraction of pulsars known in a GC does not show any
obvious pattern in the $\rho_0$--[Fe/H] plane.

\item LMXBs reside predominantly in very dense GCs, but the pulsars
appear more evenly distributed. A key question is {\em what kind\/}
of LMXBs are progenitors of GC MSPs (see \S 6).

\end{itemize}

\section{Applications of Pulsars in Globular Clusters}

\subsection{47~Tucanae}

\subsubsection{Radial Distribution}

All 17 of the 47~Tuc pulsars with a precisely known position are located
within $1\farcm2$ of the center of the cluster, even though the area of
the Parkes telescope beam that discovered them is about 100 times larger.
The deprojected spatial density is $n_p(r) \propto r^{-2}$, with none at
$r \ga 3\,r_c$ (Freire et al.\ 2001a).  While the slope of the density
profile is consistent with thermal equilibrium (assuming a dominant
stellar species of mass $\sim 1.5$\,$M_\odot$), the sharp ``edge''
in the radial extent may not be (Rasio 2000, but see also Heinke et
al.\ 2005).  In M15, by contrast, $n_p(r) \propto r^{-3}$, suggesting
a dominant species with lower mass $\sim 0.9$\,$M_\odot$ WDs, possibly
resulting from a flat IMF (Phinney 1993; Kulkarni \& Anderson 1996).

\subsubsection{Intracluster Gas and Accelerations}

Pulsars located on the ``far side'' of the GC from our point of view
are accelerated toward us/the center of the GC and if this (negative)
line-of-sight acceleration $a_l/c$ is greater than the intrinsic
(positive) $\dot P/P$ of a pulsar, the observed $\dot P < 0$.  In 47~Tuc
we see evidence that the pulsars with $\dot P < 0$ also have slightly
greater DMs.  The most straightforward interpretation of this is that at
least the central regions of the GC are permeated by a tenuous plasma:
$n_e = (0.067 \pm 0.015)$\,cm$^{-3}$.  Assuming one proton for every free
electron, the mass of this gas is $\sim 0.1$\,$M_\odot$ within 2.5\,pc
of the center (Freire et al.\ 2001b).  This is much less than the $\sim
100$\,$M_\odot$ expected to accumulate in $\sim 10^{7-8}$\,yr between
passages of the GC through the Galactic disk, and perhaps the pulsars
themselves expel most of this gas (Spergel 1991).

Also, a bound placed on $a_l$ for a given pulsar leads to a bound on
surface mass density.  Together with optical isophotes we can then
obtain a bound on mass-to-light ratios.  For instance, $M/L(r<12'') >
1.4$\,$M_\odot$/$L_\odot$ in 47~Tuc (Freire et al.\ 2003).

\subsubsection{Luminosity Function and Population}

It is not straightforward to determine the luminosity function for
the pulsars in 47~Tuc because of their large-amplitude scintillations.
Nevertheless, with careful averaging of many observations for the stronger
pulsars, and with the assumption that on average scintillation affects
all pulsars equally, Camilo et al.\ (2000) obtained an estimate for
the average flux density of 14 pulsars.  This leads to a luminosity
function $d \log N = - d \log L$ (e.g., McConnell et al.\ 2004).
{\em Assuming\/} that the minimum luminosity for 47~Tuc MSPs is $L_{1400}
\la 0.1$\,mJy\,kpc$^2$, as in the disk of the Galaxy, {\em and\/} that
the luminosity function maintains its form to these low levels, there
should be 10 times as many pulsars in the range 0.1--1\,mJy\,kpc$^2$
as there are at 1--10\,mJy\,kpc$^2$ (about 20).  This is the source
(Camilo et al.\ 2000) of the oft-quoted estimate for $\sim 200$ pulsars
in 47~Tuc (and does not even take into account a possible undercount
due to beaming effects).

However, recent radio and X-ray results seem at variance with such a
large population of pulsars.  McConnell et al.\ (2004) tried to detect
unresolved radio emission from a large number of very weak pulsars at
the center of 47~Tuc (cf.\ Fig.~1 for Ter~5).  Using their limits and
the luminosity function inferred for the higher-luminosity objects they
conclude that not many more pulsars could exist, perhaps a grand total
of 30.  From recent \textit{Chandra} observations, Heinke et al.\ (2005)
estimate that no more than 60 pulsars likely exist in 47~Tuc, regardless
of radio beaming fractions.  Both these very different arguments would
suggest, if correct, that we will not detect many more radio pulsars in
47~Tuc than the 22 known at present.  Considering the selection effects
inherent in the radio searches, however, we suggest that this question
has yet to be resolved with more sensitive searches.  In the meantime,
it seems fair to suppose that there are ``only'' 30--60 pulsars in 47~Tuc.

\subsection{Populations in Other Clusters}

Seven GCs are now known to have five or more pulsars.  The populations
of those in 47~Tuc, M5, M13, M62, and NGC~6752 appear uniform in that
they have narrow period distributions, $2 \la P \la 10$\,ms, while
those in Ter~5 and M15 display a much broader range of $P$ (Table~1).
For the newly identified large population of pulsars in Ter~5 (Ransom et al.\
2005), the luminosity distribution appears consistent with that of
47~Tuc (whether most of the diffuse radio flux shown in Fig.~1 can be
accounted for by these pulsars remains to be determined).  In any case,
as noted before, many surveys have a luminosity limit $L_{1400} \ga
10$\,mJy\,kpc$^2$, while the maximum luminosity for pulsars in 18 GCs
is $\sim 10$\,mJy\,kpc$^2$ (the exceptions are the much brighter PSRs
B1310+18, B1745--20, B1820--30A, B1821--24, and Ter~5~A).  It seems,
therefore, that in many cases at radio wavelengths we are still only
probing the tip of the iceberg.  For comparison, the total number of
pulsars present in the Galactic GC system may range from $\sim 1000$
(e.g., Heinke et al.\ 2005) to $\sim 10000$ (e.g., Kulkarni \& Anderson
1996)

\textit{Chandra} observations provide a very useful complementary picture
to the radio band (e.g., Grindlay et al.\ 2001, 2002; Grindlay, in this
volume), especially of those clusters with a small neutral hydrogen
absorbing column (since many pulsars are relatively soft X-ray sources).
It is now possible to discern in \textit{Chandra} images what must be
substantial populations of neutron stars in some clusters (e.g., Pooley
et al.\ 2003) even before we have detected them via pulsations.

\subsection{Other Applications}

Pulsars allow for a variety of other applications that we cannot discuss
in any detail in this short review.  Here we mention briefly some of
these applications.

\subsubsection{Constraining GC Dynamics}

This has been discussed extensively by Phinney (1992, 1993), but there are
many more recent examples.  For instance, the $M/L$ in the NGC~6752 core
is very high (D'Amico et al.\ 2002), while that for M62 is ``normal'' ---
and an apparent dearth of isolated MSPs in the latter (0 out of 6 total)
may hint at the dynamical state of the GC (Possenti et al.\ 2003). See
also the discussion by Lommen et al., in this volume, of the possibility
of detecting black hole binaries in GCs through MSP timing.

Other interesting applications should become possible in the near future
by measuring (or obtaining useful limits on) proper motions of some
pulsars with respect to their GC centers (e.g., Freire et al.\ 2001a).

\subsubsection{Constraining Pulsar Parameters} 

Knowing, from GC properties, the maximum expected acceleration at
a cluster center, $a_{l\,\rm max}$ (Phinney 1993), one can often
obtain a useful limit on the intrinsic $\dot P_{\rm int}$ of a pulsar,
$(\dot{P}/P)_{\rm int} < |a_{l\,\rm max}/c| + (\dot{P}/P)_{\rm obs}$,
and hence on characteristic age  $\tau_c = P/2\dot P$ and inferred
surface dipole magnetic field strength $B \propto (P \dot P)^{1/2}$.
For example, all four of M13's pulsars have $\tau_c \ga 1$\,Gyr and $B
\la 10^9$\,G (Ransom et al., in this volume), and similarly for many of
47~Tuc's pulsars (e.g., Freire et al.\ 2001a).

\subsubsection{Physical Conditions of Pulsars and Companions}

Through the study of eclipses (e.g., D'Amico et al.\ 2001), X-ray
observations of the pulsars (and possibly of some companions; e.g.,
Grindlay et al.\ 2002; Bassa et al.\ 2004; Grindlay, in this volume),
and optical observations of the companions (with five systems now clearly
detected; e.g., Edmonds et al.\ 2001, 2002; Sabbi et al.\ 2003; Sigurdsson
et al.\ 2003; Bassa et al.\ 2003; van Kerkwijk et al., in this volume),
it is possible to begin characterizing the physical conditions of the
pulsars and of their companions. For example, the detection of the WD
companion to PSR~B1620$-$26 by \textit{HST} has led to an estimate of
the age of the system, with important consequences for the origin of the
more distant, planetary companion (Sigurdsson et al.\ 2003; Sigurdsson \&
Thorsett, in this volume).

\vspace{-0.3mm}
\section{Formation and Evolution Processes}

\vspace{-1.0mm}
\subsection{Dynamical Formation Processes}

The properties of GC pulsars are quite different from those of the field
population. There is a greater proportion of single pulsars in clusters,
and the majority of the binaries have very short periods compared to field
binary pulsars. Many of these binaries have properties similar to those
of the rare eclipsing ``black widow'' pulsars seen in the Galactic disk
population (see the review by Freire, in this volume).  These systems have
extremely short orbital periods, $P_b\sim1$--10\,hr, circular orbits,
and very low-mass companions, with $m_2\simeq0.01$--0.04\,$M_\odot$.
Many of the other, ``normal'' binaries have properties more similar to
those of the disk population of low-mass binary pulsars (LMBPs), with
nearly-circular orbits, periods $P_b\sim1$--2\,d (near the short-period
end of the distribution for such binaries in the disk) and WD companions
with $m_2\sin i\simeq 0.2\,M_\odot$ (see Figs.~3 and~4).

The large inferred total population of MSPs in GCs ($\sim 50$ in 47~Tuc
alone; see \S 5.1) and the very high stellar densities in many cluster
cores ($\rho_c\sim 10^4$--$10^6\,M_\odot\,{\rm pc}^{-3}$) suggest that
dynamical interactions must play a dominant role in the formation of
these systems.  A similar conclusion is reached by considering LMXBs
in clusters, which are the likely progenitors of binary MSPs.  It was
recognized almost 30 years ago that the total number of LMXBs observed
in GCs indicates clearly a dynamical origin, with formation rates
exceeding those in field populations by several orders of magnitude
(Clark 1975). Indeed, the stellar encounter rate in a cluster core is
an excellent predictor for the presence of a bright LMXB (Pooley et al.\
2003; Jord{\' a}n et al.\ 2004).

The types of dynamical interactions involving NSs in GCs can be divided
into two categories: {\em two-body interactions\/}, which include close
tidal encounters and physical collisions, and interactions involving more
than two objects, i.e., where at least one is a binary. A particularly
important type is an {\em exchange interaction\/}, where one of the
two binary components is replaced by another star. The other star could
be a single NS, which can therefore acquire a binary companion through
this process. Alternatively, a previously formed binary MSP, or a binary
containing a non-recycled NS, could interact with another star or binary.
This can lead to a new companion for a MSP, or for a non-recycled NS,
or could release a MSP from a binary, creating a single MSP.

\subsection{Two-body Interactions}

\subsubsection{Tidal Captures}

Older scenarios based on the formation of binaries by {\em tidal
capture\/} of low-mass MS stars by NSs (Fabian, Pringle \& Rees
1975), followed by accretion and recycling of the NS during a stable
mass-transfer phase, have run into many difficulties.  First, the
formation of a long-lived binary following tidal capture is very
unlikely. This is because nonlinearities in the regime relevant to
globular clusters lead to significant energy dissipation in the MS star on
a timescale shorter than the orbital period after capture, resulting in
the rapid expansion of the star and a merger, rather than the formation
of a detached binary (Kumar \& Goodman 1996; McMillan, Taam \& McDermott
1990; Rasio \& Shapiro 1991). Moreover, the basic predictions of tidal
capture scenarios are at odds with many observations of binaries and
pulsars in clusters (Bailyn 1995; Johnston, Kulkarni \& Phinney 1992;
Shara et al.\ 1996).  It is likely that ``tidal-capture binaries''
are either never formed, or contribute negligibly to the production of
recycled pulsars (see Ivanova et al., in this volume).

The viability of tidal capture scenarios has become less relevant with
the realization in the 1990s that globular clusters contain dynamically
significant populations of {\em primordial binaries\/} (Hut et al.\
1992). Dynamical interactions involving hard primordial binaries are now
thought to provide the dominant energy production mechanism that allows
many globular clusters to remain in thermal equilibrium and avoid core
collapse over very long timescales ($\ga 10^{10}\,$yr; Gao et al.\ 1991;
McMillan \& Hut 1994; Fregeau et al.\ 2003).

\subsubsection{Stellar Collisions}

Physical {\em collisions\/} between stars (including mergers from
tidal captures) must be occurring often in dense GC cores. This is
particularly true in the presence of primordial binaries, which act as
catalysts for collisions (Bacon, Sigurdsson \& Davies 1996; Fregeau et
al.\ 2004). The products of collisions between MS stars are directly
observable in the form of {\em blue stragglers\/}.  These are objects
that appear as MS stars above the turnoff point in the color--magnitude
diagram of a cluster. All observations suggest that they must be formed
through mergers of lower-mass MS stars.  Some blue straggler masses
have been measured directly, confirming that they are more massive than
a turnoff star (Shara, Saffer \& Livio 1997; Sepinsky et al.\ 2002).
Many observations of radial profiles of blue stragglers also confirm
that they are more massive than other stars (and therefore more centrally
concentrated, as expected from mass segregation; see, e.g., Guhathakurta
et al.\ 1998; Heinke et al.\ 2003).

Collisions involving a NS have been studied using 3-D hydrodynamic
simulations (Rasio \& Shapiro 1991; Davies, Benz \& Hills 1992).  For a
collision with a MS star, the outcome is the complete destruction of the
star, and the formation of a thick, rapidly rotating envelope around
the NS. The lifetime of this envelope is highly uncertain, and it is
not clear that the NS is able to accrete enough material to be recycled
to a MSP. If it is, then this is a possible formation process for single
MSPs that does not involve the disruption of a binary (Krolik, Meiksin \&
Joss 1984). In particular, it might explain the large numbers of single,
mildly recycled MSPs in clusters with extremely high central densities,
such as M15.

\subsubsection {Collisions with Red Giants}

In contrast to NS--MS collisions, collisions of NSs with red giant
(RG) stars {\em always lead to the formation of a binary\/} (Rasio \&
Shapiro 1991). This is because the RG core always survives and ends up
in a high-eccentricity orbit around the NS.  Typically $\sim30\%$ of the
RG envelope is ejected to infinity, while most of the rest becomes bound
to the NS.  Only about $\sim 0.1\,M_\odot$ remains bound to the RG core,
which will eventually cool to a degenerate WD.  The material left bound
to the NS will attempt to form an accretion disk as it cools. The fate
of this material is again highly uncertain. It could be accreted onto
the NS and spin it up to millisecond periods (in $\sim 10^6\,$yr at
the Eddington limit), or it could be ejected, if the energy released
by accretion couples well to the gas. With an efficiency $\epsilon$,
the entire mass of gas could be ejected to infinity in as little as
$\tau_{\rm gas} \sim 10^4\,(\epsilon/0.01)^{-1}\,$yr. This short lifetime
suggests that (1) the orbit may well remain eccentric (in the absence
of an extended gaseous envelope, no tidal circularization is possible);
(2) the NS would only get mildly recycled.

Thus RG--NS collisions appear to provide a natural formation process for
eccentric LMBPs with WD companions around mildly recycled pulsars, such
as NGC~6539~A or Ter~5~J (Rasio \& Shapiro 1991; Thorsett et al.\ 1993;
Ransom et al.\ 2005). Systems with higher-mass companions, fast MSPs,
and very high eccentricities, such as NGC~1851~A (Freire et al.\ 2004;
also in this volume), are more likely the result of exchange interactions
(\S 6.3), i.e., the presently observed companion was likely acquired
later and is not the donor from which the NS was recycled. A circularized
binary MSP can also be perturbed to a higher eccentricity by a passing
star in a flyby (\S 6.3).

Note that the eccentric LMBPs found in clusters must all be formed
through dynamical processes, as there is no primordial binary evolution
channel that can produce an eccentric binary with a recycled pulsar and a
low-mass companion (cf.\ article by Kalogera et al., in this volume: the
eccentric binaries considered there all contain a {\em young\/} pulsar).

\subsubsection{Ultracompact Binaries from Collisions}

The RG--NS collisions may also play an important role in the formation
of {\em ultracompact X-ray binaries\/} (UCXBs) in clusters. These are
persistent, bright LMXBs ($L_x\sim 10^{36}$--$10^{39}\,{\rm erg}\,{\rm
s}^{-1}$) where the NS is accreting from a low-mass, degenerate companion
in a very tight orbit of period $P_b \la 1\,$hr. UCXBs may well be
dominant among the bright LMXBs observed in old globular clusters, both
Galactic (Deutsch, Margon \& Anderson 2000; van der Sluys, Verbunt \&
Pols 2004) and extragalactic (Bildsten \& Deloye 2004).  They must
connect in a fundamental way to NS recycling, as suggested by the fact
that three out of five accretion-powered millisecond X-ray pulsars known
in our Galaxy are UCXBs (Chakrabarty, in this volume).  In addition,
UCXBs may well be the progenitors of the many black-widow MSPs with
very low-mass companions observed in GCs (Rasio, Pfahl \& Rappaport 2000
and \S 6.3 below).

Several possible dynamical formation processes for UCXBs have been
discussed in the literature.  Exchange interactions between NSs and
primordial binaries provide a natural way of forming possible progenitors
of UCXBs (Davies \& Hansen 1998; Rasio et al.\ 2000; \S 6.3).  This may
well dominate the formation rate when integrated over the entire GC
dynamical history. However, it is unlikely to be significant for bright
UCXBs observed {\em today\/} in clusters. This is because the progenitors
must be intermediate-mass binaries, with the NS companion massive enough
for the initial mass transfer to become dynamically unstable, leading to
common-envelope (CE) evolution and significant orbital decay. Instead,
all MS stars remaining today in an old GC (with masses below the turn-off
mass $m_{\rm to}\simeq 0.8\,M_\odot$) have masses low enough to lead
to {\em stable\/} mass transfer (and orbits that expand during mass
transfer, leading to LMXBs with wide periods and non-degenerate donors).
Alternatively, some binaries with stable mass transfer could evolve to
ultra-short periods under the influence of magnetic braking (Pylyser
\& Savonije 1988; Podsiadlowski, Rappaport \& Pfahl 2002). However,
producing UCXBs through this type of evolution requires very careful
tuning of initial conditions, and it is therefore very unlikely to
explain most sources in GCs (van der Sluys et al.\ 2004).

Verbunt (1987) first proposed that RG--NS collisions could lead to
UCXB formation. In the original scenario, the collision was assumed to
lead directly to a CE system in which the NS and RG core would quickly
inspiral. However, RG--NS collisions that occur now in old globular
clusters (where RGs have low masses, close to $m_{\rm to}$) do {\em
not\/} lead to CE evolution. Instead, as noted above, the RG envelope is
promptly disrupted, leaving behind an eccentric NS--WD binary (Rasio \&
Shapiro 1991).  Nevertheless, if the post-collision NS--WD binaries can
retain their high eccentricities, then many of these systems could decay
through gravitational-wave emission all the way to contact and still
become UCXBs (Davies et al.\ 1992; Ivanova et al.\ 2005).

\subsection{Interactions with Binaries}

\subsubsection{Binary Flybys}

As noted in \S 3.3, the eccentricity of a binary can be perturbed
significantly during a close flyby of another object in the cluster.
Secular perturbations of the eccentricity scale as a power-law in the
distance of closest approach, and can therefore be significant even
in fairly distant interactions (Heggie \& Rasio 1996).  In contrast,
semi-major axis (energy) perturbations decay {\em exponentially\/}
with distance of closest approach (Heggie 1975).  Since the intrinsic
eccentricities of tidally circularized binary MSPs can be extremely small
(down to $\la 10^{-6}$ for Galactic disk binaries; see right panel of
Fig.~4), the currently measured eccentricities of LMBPs in clusters
provide a sensitive record of past dynamical interactions (Phinney 1992;
Rasio \& Heggie 1995).

However, inducing a significant eccentricity (say $e > 0.1$) in an
initially circular binary requires a very close flyby, because there is no
perturbation to the eccentricity of an initially circular orbit to lowest
order in secular perturbations (Heggie \& Rasio 1996). As a consequence,
the most likely result from a flyby is a nonzero but small eccentricity
($e \la 0.1$) for an otherwise usual LMBP with a reasonably wide orbit
($P_b \ga 1\,$d).  For example, the $\sim 0.1$ eccentricity of M5~B
($P_b\simeq 7\,$d) is entirely consistent with having been induced by
close interactions with passing stars in the cluster (Rasio \& Heggie
1995).  It is likely that many binaries with $e\sim 10^{-5}$--$10^{-1}$
appearing well above the theoretical eccentricity--period relation in
Fig.~4 (right) have been similarly perturbed through flybys.

\subsubsection{Exchange Interactions}

Single NSs retained in a dense GC can easily acquire binary companions
through {\em exchange interactions\/} with primordial binaries. Because
of its large cross section, this process tends to dominate over two-body
interactions, at least for sufficiently high binary fractions (Heggie,
Hut \& McMillan 1996; Leonard 1989; Sigurdsson \& Phinney 1993).

In contrast to tidal capture, exchange interactions with hard primordial
binaries (with semi-major axes $a\sim0.1$--1\,AU) can form naturally
the wide LMBPs seen in some low-density globular clusters (such as PSR
B1310$+$18, with $P_b=256\,$d, in M53, which has the lowest central
density, $\rho_c\sim10^3\,M_\odot\,{\rm pc}^{-3}$, of any globular
cluster with detected radio pulsars; see right panel of Fig.~6).  When the
newly acquired companion star, of mass $\la1\,M_\odot$, evolves up the
giant branch, the orbit circularizes and a period of {\em stable\/} mass
transfer begins, during which the NS is recycled (see, e.g., Rappaport et
al.\ 1995).  The resulting MSP--WD binaries have orbital periods in the
range $P_b\sim1$--$10^3\,$d (see Willems \& Kolb, in this volume). Very
wide binaries formed in this way will interact again easily, possibly
releasing the NS as a single MSP.

However, this scenario cannot explain the formation of recycled pulsars
in binaries with periods shorter than $\sim1\,$d. To obtain such short
periods, the initial primordial binary must be extremely hard, with
$a \la 0.01\,$AU, but then the recoil velocity of the system following
the exchange interaction would almost certainly exceed the escape speed
from the shallow cluster potential (e.g., $v_e\simeq 60\,{\rm km}\,{\rm
s}^{-1}$ for 47~Tuc).

\subsubsection{Intermediate-Mass Binaries}

One can get around this problem by considering more carefully
the stability of mass transfer in binaries formed through exchange
interactions.  While all MS stars in the cluster {\em today\/} have masses
$\la1\,M_\odot$, the rate of exchange interactions should have peaked
at a time when significantly more massive MS stars were still present.
Indeed, the NSs and the most massive primordial binaries will undergo
mass segregation and concentrate in the cluster core on a timescale
comparable to the initial half-mass relaxation time $t_{rh}$. For
typical dense globular clusters, we expect $t_{rh}\sim 10^9\,$yr, which
is comparable to the MS lifetime of a $\sim 2$--$3\,M_\odot$ star.
The exchange interactions will then lead to the formation of {\em
intermediate-mass binaries\/} (Davies \& Hansen 1998).  Among LMBPs
in the Galactic disk, at least one system (PSR J2051$-$0827) is likely
to have had an intermediate-mass binary progenitor, given its very low
transverse velocity (Stappers et al.\ 1998).

If the majority of NSs in a cluster core acquired MS companions with
masses up to $\sim3\,M_\odot$, a very different type of evolution could
result.  Indeed, in this case, when the MS companion evolves and fills
its Roche lobe, the mass transfer for many systems (depending on the
mass ratio and evolutionary state of the donor star) is {\em dynamically
unstable\/} and leads to a CE phase (see, e.g., Taam \& Sandquist 2000).
The emerging binary will have a low-mass WD in a short-period, circular
orbit around the NS. Some of these tight NS--WD binaries will decay
to contact through gravitational wave emission, forming UCXBs, and,
ultimately, black-widow pulsars (Rasio et al.\ 2000).

Simple Monte Carlo simulations of this process can produce a variety
of short-period binaries with properties that agree well with those
of observed MSPs in 47~Tuc (Rasio 2003).  The results also predict
the existence of a large number of binary MSPs with companion
masses $m_2\simeq0.03$--0.05\,$M_\odot$ and orbital periods as
short as $\sim15\,$min (descendants of UCXBs) that may have so far
escaped detection.  Future observations using more sophisticated
acceleration-search techniques or shorter integration times may be able
to detect them.

\subsubsection{Multiple Interactions}

Exchange interactions can also involve previously formed LMBPs.  With the
MSP liberated from the binary in which it was formed, this provides the
simplest mechanism for producing single MSPs in clusters (and explaining
their higher incidence in GCs than in the field).  In addition, exchange
interactions can lead to the replacement of the original MSP companion by
a new, unexpected companion. The resulting orbit can be highly eccentric,
and the system can be ejected from the cluster core so that, if the
interaction was recent enough (compared to the relaxation time near the
new orbit's apocenter), the system may now be observed with an unusually
large offset from the center. The most exotic products might be NS--NS
systems, such as M15~C (Phinney \& Sigurdsson 1991), or MSP--black hole
binaries (Sigurdsson 2003).  More commonly, the new companion would be a
MS star, explaining systems such as PSR J1740$-$5340 in the outskirts of
NGC~6397. If this MS star later evolves and attempts (a second episode
of) mass transfer onto the MSP, unusual binary evolution may ensue,
with the MSP wind preventing accretion onto the NS, and all mass from
the companion leaving the system (Nelson, in this volume). This type of
evolution has been proposed as a way of forming some of the black-widow
binaries (King, Davies \& Beer 2003).

\bigskip

\acknowledgements We are very grateful to the Aspen Center for Physics
for hospitality and financial support.  This work was supported in part
by a \textit{Chandra} Theory Grant, NASA Grant NAG5-12044, NSF Grants
AST-02-06276 and AST-02-05853, and a NRAO Travel Grant.  FC thanks his
many collaborators, especially S.\ Ransom, P.\ Freire, D.\ Lorimer,
and A.\ Possenti.  FAR thanks L.\ Bildsten, N.\ Ivanova, V.\ Kalogera,
and J.\ Lombardi for helpful discussions.

\begin{table}
\caption{Parameters for 100 pulsars known in 24 globular clusters.}
\footnotesize
\label{tab:psr_parms}
\begin{center}
\begin{tabular}{lllllllll}
\hline
\hline
\multicolumn{1}{l}{Pulsar}          &
\multicolumn{1}{l}{$P$}             &
\multicolumn{1}{l}{\rule{0cm}{3.5mm}$\dot P$}        &
\multicolumn{1}{l}{$r$}             &
\multicolumn{1}{l}{$P_b$}           &
\multicolumn{1}{l}{$x$}             &
\multicolumn{1}{l}{$e$}             &
\multicolumn{1}{l}{$m_2$}           &
\multicolumn{1}{l}{Ref}             \\
\multicolumn{1}{l}{}                &
\multicolumn{1}{l}{(ms)}            &
\multicolumn{1}{l}{($10^{-20}$)}    &
\multicolumn{1}{l}{($r_c$)}         &
\multicolumn{1}{l}{(d)}             &
\multicolumn{1}{l}{(s)}             &
\multicolumn{1}{l}{}                &
\multicolumn{1}{l}{($M_\odot$)}     &
\multicolumn{1}{l}{}\vspace{1mm}    \\
\hline
\hline
\multicolumn{9}{c}{\rule{0mm}{3.5mm}47~Tuc ($r_c$=0.40, $r_h$=2.79, $c$=2.03, $\rho_0$=4.81, $t_c$=7.96, $t_h$=9.48, [Fe/H]=--0.76, $R$=4.5)} \vspace{1mm} \\
\hline
\vspace{-3mm}\\
J0023$-$7204C & 5.756 & $-$4.98 & 3.02  &       &       &            & & 1 \\
J0024$-$7204D & 5.357 & $-$0.34 & 1.70  &       &       &            & & 1 \\
J0024$-$7205E & 3.536 & +9.85   & 1.62  & 2.256 & 1.981 & 0.000315   & 0.15 & 1 \\
J0024$-$7204F & 2.623 & +6.45   & 0.47  &       &       &            & & 1 \\
J0024$-$7204G & 4.040 & $-$4.21 & 0.72  &       &       &            & & 1 \\
J0024$-$7204H & 3.210 & $-$0.18 & 1.92  & 2.357 & 2.152 & 0.07056    & 0.16 & 1 \\
J0024$-$7204I & 3.484 & $-$4.58 & 0.72  & 0.229 & 0.038 & $<0.0004$  & 0.013 & 1 \\
J0023$-$7203J & 2.100 & $-$0.97 & 2.50 & 0.120$^e$ & 0.040 & $<0.00004$ & 0.021 & 1 \\
J0024$-$7204L & 4.346 & $-$12.2 & 0.35  &       &       &            & & 1 \\
J0023$-$7205M & 3.676 & $-$3.84 & 2.62  &       &       &            & & 1 \\
J0024$-$7204N & 3.053 & $-$2.18 & 1.22  &       &       &            & & 1 \\
J0024$-$7204O & 2.643 & +3.03 & 0.01 & 0.135$^e$ & 0.045 & $<0.00016$ & 0.022 & 1 \\
            P & 3.643 &         &       & 0.147 & 0.038 &         & 0.017 & 2 \\
J0024$-$7204Q & 4.033 & +3.40   & 2.45  & 1.189 & 1.462 & 0.00008 & 0.17  & 1 \\
            R & 3.480 &         &       & 0.066$^e$ & 0.033 &     & 0.026 & 2 \\
J0024$-$7204S & 2.830 & $-$12.0 & 0.47  & 1.201 & 0.766 & 0.00039 & 0.088 & 1 \\
J0024$-$7204T & 7.588 & +29.3   & 0.85  & 1.126 & 1.338 & 0.0004  & 0.16  & 1 \\
J0024$-$7203U & 4.342 & +9.52   & 2.35  & 0.429 & 0.526 & 0.00014 & 0.12  & 1 \\
            V & 4.810 &         &       & 0.2$^e$   & 0.8   &     & 0.34  & 2 \\
J0024$-$7204W & 2.352 &         & 0.20  & 0.133$^e$ & 0.243 &     & 0.12  & 3 \\
            X & 4.771 &         &       & ?     &       &         &       & 4 \\
            Y & 2.196 &         &       & 0.521 & 0.671 &         & 0.13  & 4 \\
\hline
\hline
\multicolumn{9}{c}{\rule{0mm}{3.5mm}NGC~1851 ($r_c$=0.06, $r_h$=0.52, $c$=2.32, $\rho_0$=5.32, $t_c$=6.98, $t_h$=8.85, [Fe/H]=--1.22, $R$=12.1)} \vspace{1mm} \\
\hline
\vspace{-3mm}\\
J0514$-$4002 & 4.990 & & 1.72 & 18.785 & 36 & 0.88 & 0.89 & 5 \\
\hline
\hline
\multicolumn{9}{c}{\rule{0mm}{3.5mm}M53 ($r_c$=0.36, $r_h$=1.11, $c$=1.78, $\rho_0$=3.05, $t_c$=8.76, $t_h$=9.66, [Fe/H]=--1.99, $R$=17.8)} \vspace{1mm} \\
\hline
\vspace{-3mm}\\
B1310+18 & 33.163 & & & 255 & 84 & $<0.01$ & 0.29 & 6 \\
\hline
\hline
\multicolumn{9}{c}{\rule{0mm}{3.5mm}M3 ($r_c$=0.55, $r_h$=1.12, $c$=1.84, $\rho_0$=3.51, $t_c$=8.84, $t_h$=9.35, [Fe/H]=--1.57, $R$=10.4)} \vspace{1mm} \\
\hline
\vspace{-3mm}\\
A & 2.545 & & & ?    &     & & & 7 \\
B & 2.389 & & & 1.42 & 1.9 & & 0.20 & 7 \\
C & 2.166 & & & ?    &     & & & 7 \\
D & 5.443 & & & ?    &     & & & 7 \\
\hline
\hline
\multicolumn{9}{c}{\rule{0mm}{3.5mm}M5 ($r_c$=0.42, $r_h$=2.11, $c$=1.83, $\rho_0$=3.91, $t_c$=8.26, $t_h$=9.53, [Fe/H]=--1.27, $R$=7.5)} \vspace{1mm} \\
\hline
\vspace{-3mm}\\
B1516+02A & 5.553 & +4.12  & 1.19 &       &       &        & & 8 \\
B1516+02B & 7.946 & $-$0.3 & 0.71 & 6.858 & 3.048 & 0.1378 & 0.11 & 8 \\
        C & 2.484 &        &      & 0.087$^e$ & 0.057 &    & 0.037 & 7 \\
        D & 2.988 &        &      & 1.22 & 1.6    &        & 0.19 & 7 \\
        E & 3.182 &        &      & 1.10 & 1.2    &        & 0.15 & 7 \\
\hline
\hline
\multicolumn{9}{c}{\rule{0mm}{3.5mm}M4 ($r_c$=0.83, $r_h$=3.65, $c$=1.59, $\rho_0$=3.82, $t_c$=7.57, $t_h$=8.82, [Fe/H]=--1.20, $R$=2.2)} \vspace{1mm} \\
\hline
\vspace{-3mm}\\
B1620$-$26 & 11.075 & $-$5.46 & 0.92 & 191.442 & 64.809 & 0.025315 & 0.27 & 9 \\
\hline
\hline
\multicolumn{9}{c}{\rule{0mm}{3.5mm}M13 ($r_c$=0.78, $r_h$=1.49, $c$=1.51, $\rho_0$=3.33, $t_c$=8.80, $t_h$=9.30, [Fe/H]=--1.54, $R$=7.7)} \vspace{1mm} \\
\hline
\vspace{-3mm}\\
B1639+36A & 10.377 & $<4.5$  &  &       &      &          & & 6 \\
B1639+36B &  3.528 &         &  & 1.259 & 1.38 & $<0.001$ & 0.15 & 10 \\
        C &  3.722 &         &  &       &      &          & & 7 \\
        D &  3.118 &         &  & 0.591 & 0.92 &          & 0.17 & 7 \\
        E &  2.487 &         &  & 0.213 & 0.17 &          & 0.061 & 7 \\
\hline
\hline
\end{tabular}
\end{center}
\end{table}

\addtocounter{table}{-1}
\begin{table}
\caption{(continued).}
\vspace{-2.0mm}
\footnotesize
\begin{center}
\begin{tabular}{lllllllll}
\hline
\hline
\multicolumn{1}{l}{Pulsar}          &
\multicolumn{1}{l}{$P$}             &
\multicolumn{1}{l}{\rule{0cm}{3.5mm}$\dot P$}        &
\multicolumn{1}{l}{$r$}             &
\multicolumn{1}{l}{$P_b$}           &
\multicolumn{1}{l}{$x$}             &
\multicolumn{1}{l}{$e$}             &
\multicolumn{1}{l}{$m_2$}           &
\multicolumn{1}{l}{Ref}             \\
\multicolumn{1}{l}{}                &
\multicolumn{1}{l}{(ms)}            &
\multicolumn{1}{l}{($10^{-20}$)}    &
\multicolumn{1}{l}{($r_c$)}         &
\multicolumn{1}{l}{(d)}             &
\multicolumn{1}{l}{(s)}             &
\multicolumn{1}{l}{}                &
\multicolumn{1}{l}{($M_\odot$)}     &
\multicolumn{1}{l}{}\vspace{1mm}    \\
\hline
\hline
\multicolumn{9}{c}{\rule{0mm}{3.5mm}M62 ($r_c$=0.18, $r_h$=1.23, $c$=1.70c, $\rho_0$=5.14, $t_c$=7.64, $t_h$=9.19, [Fe/H]=--1.29, $R$=6.9)} \vspace{0.5mm} \\
\hline
\vspace{-3mm}\\
J1701$-$3006A & 5.241 & $-13.19$ & 1.77 & 3.805     & 3.483 & $<0.000004$& 0.19 & 11 \\
J1701$-$3006B & 3.593 & $-34.97$ & 0.01 & 0.144$^e$ & 0.252 & $<0.00007$ & 0.12 & 11 \\
J1701$-$3006C & 3.806 & $-3.18$  & 0.97 & 0.215     & 0.192 & $<0.00006$ & 0.069 & 11 \\
            D & 3.418 &          & & 1.12  & 0.98  &               & 0.12 & 12 \\
            E & 3.234 &          & & 0.16$^e$  & 0.07  &           & 0.030 & 12 \\
            F & 2.295 &          & & 0.20  & 0.05  &               & 0.018 & 12 \\
\hline
\hline
\multicolumn{9}{c}{\rule{0mm}{3.5mm}NGC~6342 ($r_c$=0.05, $r_h$=0.88, $c$=2.50c, $\rho_0$=4.77, $t_c$=6.09, $t_h$=8.66, [Fe/H]=--0.65, $R$=8.6)} \vspace{0.5mm} \\
\hline
\vspace{-3mm}\\
B1718$-$19$^*$ & 1004.03 & +150000 & 46.0 & 0.258$^e$ & 0.352 & $<0.005$ & 0.11 & 13 \\
\hline
\hline
\multicolumn{9}{c}{\rule{0mm}{3.5mm}NGC~6397 ($r_c$=0.05, $r_h$=2.33, $c$=2.50c, $\rho_0$=5.68, $t_c$=4.90, $t_h$=8.46, [Fe/H]=--1.95, $R$=2.3)} \vspace{0.5mm} \\
\hline
\vspace{-3mm}\\
J1740$-$5340 & 3.650 & +16 & 18.3 & 1.354$^e$ & 1.652 & $<0.0001$ & 0.18 & 14 \\
\hline
\hline
\multicolumn{9}{c}{\rule{0mm}{3.5mm}NGC~6440 ($r_c$=0.13, $r_h$=0.58, $c$=1.70, $\rho_0$=5.28, $t_c$=7.54, $t_h$=8.76, [Fe/H]=--0.34, $R$=8.4)} \vspace{0.5mm} \\
\hline
\vspace{-3mm}\\
B1745$-$20 & 288.602 & +40000 & 0.76 & & & & & 15 \\
\hline
\hline
\multicolumn{9}{c}{\rule{0mm}{3.5mm}Terzan~5 ($r_c$=0.18, $r_h$=0.83, $c$=1.87, $\rho_0$=5.06, $t_c$=8.16, $t_h$=8.97, [Fe/H]=0.00, $R$=10.3)} \vspace{0.5mm} \\
\hline
\vspace{-3mm}\\
J1748$-$2446A & 11.563 & $-3.4$ & 2.77 &  0.075$^e$ & 0.119 & $< 0.0012$ & 0.087 & 16 \\
J1748$-$2446C &  8.436 & $-60$  & 0.94 &            &       &       & & 16 \\
D        &  4.713 &        &    &            &       &            & & 17 \\
E        &  2.197 &        &    & 60.06      & 23.6  & $\sim$0.02 & 0.21 & 17 \\
F        &  5.540 &        &    &            &       &            & & 17 \\
G        & 21.671 &        &    &            &       &            & & 17 \\
H        &  4.925 &        &    &            &       &            & & 17 \\
I        &  9.570 &        &    &  1.328     & 1.818 & 0.428      & 0.20 & 17 \\
J        & 80.337 &        &    &  1.102     & 2.454 & 0.350      & 0.33 & 17 \\
K        &  2.969 &        &    &            &       &            & & 17 \\
L        &  2.244 &        &    &            &       &            & & 17 \\
M        &  3.569 &        &    &  0.443     & 0.596 &            & 0.13 & 17 \\
N        &  8.666 &        &    &  0.385     & 1.619 & 0.000045   & 0.46 & 17 \\
O        &  1.676 &        &    &  0.259$^e$ & 0.112 &            & 0.035 & 17 \\
P        &  1.728 &        &    &  0.362$^e$ & 1.272 &            & 0.36 & 17 \\
Q        &  2.812 &        &    & $>$1?      &       &            & & 17 \\
R        &  5.028 &        &    &            &       &            & & 17 \\
S        &  6.116 &        &    &            &       &            & & 17 \\
T        &  7.084 &        &    &            &       &            & & 17 \\
U        &  3.289 &        &    & $>$1?      &       &            & & 17 \\
V        &  2.072 &        &    &  0.503     & 0.567 &            & 0.11 & 17 \\
W        &  4.205 &        &    &  4.877     & 5.869 & 0.015      & 0.29 & 17 \\
X        &  2.999 &        &    & $>$1?      &       &            & & 17 \\
\hline
\hline
\multicolumn{9}{c}{\rule{0mm}{3.5mm}NGC~6441 ($r_c$=0.11, $r_h$=0.64, $c$=1.85, $\rho_0$=5.25, $t_c$=7.77, $t_h$=9.19, [Fe/H]=--0.53, $R$=11.7)} \vspace{0.5mm} \\
\hline
\vspace{-3mm}\\
J1750$-$37 & 111.609 & & & 17.3 & 24.4 & 0.71 & 0.57 & 18 \\
\hline
\hline
\multicolumn{9}{c}{\rule{0mm}{3.5mm}NGC~6539 ($r_c$=0.54, $r_h$=1.67, $c$=1.60, $\rho_0$=3.62, $t_c$=8.60, $t_h$=9.37, [Fe/H]=--0.66, $R$=8.4)} \vspace{0.5mm} \\
\hline
\vspace{-3mm}\\
B1802$-$07 & 23.100 & +47 & 0.46 & 2.616 & 3.920 & 0.212 & 0.29 & 19 \\
\hline
\hline
\multicolumn{9}{c}{\rule{0mm}{3.5mm}NGC~6522 ($r_c$=0.05, $r_h$=1.04, $c$=2.50c, $\rho_0$=5.31, $t_c$=6.32, $t_h$=8.90, [Fe/H]=--1.44, $R$=7.8)} \vspace{0.5mm} \\
\hline
\vspace{-3mm}\\
J1803$-$30    & 7.101 & & & & & & & 18 \\
\hline
\hline
\multicolumn{9}{c}{\rule{0mm}{3.5mm}NGC~6544 ($r_c$=0.05, $r_h$=1.77, $c$=1.63c, $\rho_0$=5.73, $t_c$=5.09, $t_h$=8.40, [Fe/H]=--1.56, $R$=2.7)} \vspace{0.5mm} \\
\hline
\vspace{-3mm}\\
A & 3.059 & & & 0.071 & 0.012 & & 0.0089 & 20 \\
B & 4.186 & & & ?     &       & & & 12 \\
\hline
\hline
\end{tabular}
\end{center}
\end{table}

\addtocounter{table}{-1}
\begin{table}
\caption{(continued).}
\footnotesize
\begin{center}
\begin{tabular}{lllllllll}
\hline
\hline
\multicolumn{1}{l}{Pulsar}          &
\multicolumn{1}{l}{$P$}             &
\multicolumn{1}{l}{\rule{0cm}{3.5mm}$\dot P$}        &
\multicolumn{1}{l}{$r$}             &
\multicolumn{1}{l}{$P_b$}           &
\multicolumn{1}{l}{$x$}             &
\multicolumn{1}{l}{$e$}             &
\multicolumn{1}{l}{$m_2$}           &
\multicolumn{1}{l}{Ref}             \\
\multicolumn{1}{l}{}                &
\multicolumn{1}{l}{(ms)}            &
\multicolumn{1}{l}{($10^{-20}$)}    &
\multicolumn{1}{l}{($r_c$)}         &
\multicolumn{1}{l}{(d)}             &
\multicolumn{1}{l}{(s)}             &
\multicolumn{1}{l}{}                &
\multicolumn{1}{l}{($M_\odot$)}     &
\multicolumn{1}{l}{}\vspace{1mm}    \\
\hline
\hline
\multicolumn{9}{c}{\rule{0mm}{3.5mm}NGC~6624 ($r_c$=0.06, $r_h$=0.82, $c$=2.50c, $\rho_0$=5.25, $t_c$=6.61, $t_h$=8.73, [Fe/H]=--0.44, $R$=7.9)} \vspace{0.5mm} \\
\hline
\vspace{-3mm}\\
B1820$-$30A &   5.440 & +338  & 0.83 & & & & & 21 \\
B1820$-$30B & 378.596 & +3150 & 3.83 & & & & & 21 \\
          C & 405.9   &       &      & & & & & 12 \\
\hline
\hline
\multicolumn{9}{c}{\rule{0mm}{3.5mm}M28 ($r_c$=0.24, $r_h$=1.56, $c$=1.67, $\rho_0$=4.73, $t_c$=7.58, $t_h$=9.04, [Fe/H]=--1.45, $R$=5.6)} \vspace{0.5mm} \\
\hline
\vspace{-3mm}\\
B1821$-$24 & 3.054 & +161 & 0.09 & & & & & 22 \\
\hline
\hline
\multicolumn{9}{c}{\rule{0mm}{3.5mm}NGC~6749 ($r_c$=0.77, $r_h$=1.10, $c$=0.83, $\rho_0$=3.33, $t_c$=8.90, $t_h$=8.79, [Fe/H]=--1.60, $R$=7.9)} \vspace{0.5mm} \\
\hline
\vspace{-3mm}\\
A & 3.193 & & & & & & & 7 \\ 
B & 4.968 & & & & & & & 7 \\ 
\hline
\hline
\multicolumn{9}{c}{\rule{0mm}{3.5mm}NGC~6752 ($r_c$=0.17, $r_h$=2.34, $c$=2.50c, $\rho_0$=4.91, $t_c$=6.83, $t_h$=9.01, [Fe/H]=--1.56, $R$=4.0)} \vspace{0.5mm} \\
\hline
\vspace{-3mm}\\
J1911$-$5958A & 3.266 & +0.30 & 37.5 & 0.837 & 1.206 & $<0.00001$ & 0.18 & 23 \\
J1910$-$5959B & 8.357 & $-$79 & 0.58 & & & & & 23 \\
J1911$-$6000C & 5.277 & +0.2  & 15.8 & & & & & 23 \\
J1910$-$5959D & 9.035 & +96   & 1.11 & & & & & 23 \\
J1910$-$5959E & 4.571 & $-$43 & 0.76 & & & & & 23 \\
\hline
\hline
\multicolumn{9}{c}{\rule{0mm}{3.5mm}NGC~6760 ($r_c$=0.33, $r_h$=2.18, $c$=1.59, $\rho_0$=3.84, $t_c$=7.94, $t_h$=9.39, [Fe/H]=--0.52, $R$=7.4)} \vspace{0.5mm} \\
\hline
\vspace{-3mm}\\
J1911+0102A & 3.618 & $-0.65$ & 1.27 & 0.140 & 0.037 & $<0.00013$ & 0.017 & 24 \\
J1911+0101B & 5.384 & $-0.2$  & 0.36 &       &       &            & & 24 \\
\hline
\hline
\multicolumn{9}{c}{\rule{0mm}{3.5mm}M71 ($r_c$=0.63, $r_h$=1.65, $c$=1.15, $\rho_0$=3.04, $t_c$=7.65, $t_h$=8.43, [Fe/H]=--0.73, $R$=4.0)} \vspace{0.5mm} \\
\hline
\vspace{-3mm}\\
A & 4.888 & & & 0.176$^e$ & 0.078 & & 0.032 & 7 \\
\hline
\hline
\multicolumn{9}{c}{\rule{0mm}{3.5mm}M15 ($r_c$=0.07, $r_h$=1.06, $c$=2.50c, $\rho_0$=5.38, $t_c$=7.02, $t_h$=9.35, [Fe/H]=--2.26, $R$=10.3)} \vspace{0.5mm} \\
\hline
\vspace{-3mm}\\
B2127+11A &  110.664 & $-$2107 & 0.25 & & & & & 10 \\
B2127+11B &   56.133 &    +956 & 1.12 & & & & & 10 \\
B2127+11C &   30.529 &    +499 & 13.4 & 0.335 & 2.518 & 0.681 & 0.92 & 10 \\
B2127+11D &    4.802 &  $-$107 & 0.27 & & & & & 10 \\
B2127+11E &    4.651 &     +17 & 1.92 & & & & & 10 \\
B2127+11F &    4.027 &      +3 & 3.98 & & & & & 10 \\
B2127+11G &   37.660 &    +195 & 1.51 & & & & & 10 \\
B2127+11H &    6.743 &      +2 & 0.54 & & & & & 10 \\
\hline
\hline
\multicolumn{9}{c}{\rule{0mm}{3.5mm}M30 ($r_c$=0.06, $r_h$=1.15, $c$=2.50c, $\rho_0$=5.04, $t_c$=6.38, $t_h$=8.95, [Fe/H]=--2.12, $R$=8.0)} \vspace{0.5mm} \\
\hline
\vspace{-3mm}\\
A & 11.019 & $-5.18$ & 1.11 & 0.173$^e$  & 0.234  & $<0.00012$ & 0.10 & 25 \\
B & 13.0   &         &      & $>$0.8 & $>0.1$ & $>0.52$        & & 25 \\
\hline
\hline
\end{tabular}
\end{center}
\vspace{-4.4mm}
\noindent {\em Cluster parameters\/} (Harris 1996): $r_c$ (core radius
in arcmin), $r_h$ (half-mass radius in arcmin), $c = \log (r_t/r_c)$
(central concentration where $r_t$ is tidal radius and a ``c'' denotes
core-collapsed cluster), $\rho_0$ ($\log$ of central luminosity density
in $L_\odot$\,pc$^{-3}$), $t_c$ ($\log$ of core relaxation time in yr),
$t_h$ ($\log$ of relaxation time at $r_h$ in yr), [Fe/H] (metallicity),
$R$ (distance from Sun in kpc).\\
\noindent {\em Pulsar parameters\/}: $P$ (period); where measured,
observed period derivative ($\dot P$) and angular positional offset
from GC center ($r$); for binaries: $P_b$ (orbital period; binaries
with uncertain $P_b$ are indicated by ``?''; $^e$ indicates radio
eclipses), $x$ (projected semi-major axis light travel time), $e$
(eccentricity), $m_2$ (minimum companion mass assuming a pulsar mass
$m_1=1.35\,M_\odot$).\\
\noindent $^*$ Membership in NGC~6342 for PSR~B1718--19 is not certain
(see \S 3.4).\\
\noindent {\em References\/}: 
1 (Freire et al.\ 2003);
2 (Camilo et al.\ 2000);
3 (Edmonds et al.\ 2002);
4 (Lorimer et al.\ 2003);
5 (Freire et al.\ 2004);
6 (Kulkarni et al.\ 1991);
7 (Ransom et al., this volume);
8 (Anderson et al.\ 1997);
9 (Sigurdsson et al.\ 2003);
10 (Anderson 1993);
11 (Possenti et al.\ 2003);
12 (Chandler 2003);
13 (van Kerkwijk et al.\ 2000);
14 (D'Amico et al.\ 2001);
15 (Lyne et al.\ 1996);
16 (Lyne et al.\ 2000);
17 (Ransom et al.\ 2005);
18 (Possenti, this volume);
19 (Thorsett et al.\ 1993);
20 (Ransom et al.\ 2001);
21 (Biggs et al.\ 1994);
22 (Cognard et al.\ 1996);
23 (D'Amico et al.\ 2002);
24 (Freire et al.\ 2005);
25 (Ransom et al.\ 2004).
\end{table}


\clearpage

\begin{references}

\reference Alpar, M.~A., Cheng, A.~F., Ruderman, M.~A., \& Shaham,
J. 1982, Nature, 300, 728

\reference Anderson, S.~B. 1993, PhD thesis, California Institute of
Technology

\reference Anderson, S.~B., Wolszczan, A., Kulkarni, S.~R., \& Prince,
T.~A. 1997, \apj, 482, 870

\reference Backer, D.~C., Kulkarni, S.~R., Heiles, C., Davis, M.~M., \&
Goss, W.~M. 1982, Nature, 300, 615

\reference Bacon, D., Sigurdsson, S., \& Davies, M.~B. 1996, MNRAS,
281, 830

\reference Bailyn, C.~D. 1995, ARAA, 33, 133

\reference Bassa, C.~G., Verbunt, F., van Kerkwijk, M.~H., \& Homer,
L. 2003, \aap, 409, L31

\reference Bassa, C., Pooley, D., Homer, L., Verbunt, F., Gaensler,
B.~M., Lewin, W.~H.~G., Anderson, S.~F., Margon, B., Kaspi, V.~M.,
\& van der Klis, M. 2004, \apj, 609, 755

\reference Biggs, J.~D., Bailes, M., Lyne, A.~G., Goss, W.~M., \&
Fruchter, A.~S. 1994, \mnras, 267, 125

\reference Bildsten, L., \& Deloye, C.~J. 2004, \apj, 607, L119

\reference Camilo, F., Lorimer, D.~R., Freire, P., Lyne, A.~G., \&
Manchester, R.~N. 2000, \apj, 535, 975

\reference Chandler, A.~M. 2003, PhD thesis, California Institute of
Technology

\reference Clark, G.~W. 1975, \apj, 199, L143

\reference Cognard, I., Bourgois, G., Lestrade, J.-F., Biraud, F., Aubry,
D., Darchy, B., \& Drouhin, J.-P. 1996, A\&A, 311, 179

\reference D'Amico, N., Possenti, A., Manchester, R.~N., Sarkissian,
J., Lyne, A.~G.,\& Camilo, F. 2001, \apj, 561, L89

\reference D'Amico, N., Possenti, A., Fici, L., Manchester, R.~N., Lyne,
A.~G., \& Camilo, F. 2002, \apj, 570, L89

\reference Davies, M.~B., \& Hansen, B.~M.~S. 1998, MNRAS, 301, 15

\reference Davies, M.~B., Benz, W., \& Hills, J.~G. 1992, \apj, 401, 246

\reference Deutsch, E.~W., Margon, B., \& Anderson, S.~F. 2000, \apj,
530, L21

\reference Edmonds, P.~D., Gilliland, R.~L., Heinke, C.~O., Grindlay,
J.~E., \& Camilo, F. 2001, \apj, 557, L57

\reference Edmonds, P.~D., Gilliland, R.~L., Camilo, F., Heinke, C.~O.,
\& Grindlay, J.~E. 2002, \apj, 579, 741

\reference Fabian, A.~C., Pringle, J.~E., \& Rees, M.~J. 1975, MNRAS,
172, P15

\reference Ferraro, F.~R., Sabbi, E., Gratton, R., Possenti, A., D'Amico,
N., Bragaglia, A., \& Camilo, F. 2003, \apj, 584, L13

\reference Fregeau, J.~M., Cheung, P., Portegies-Zwart, S.~F., \& Rasio,
F.~A. 2004, MNRAS, 352, 1

\reference Fregeau, J.~M., G{\"u}rkan, A., Joshi, K.~J., \& Rasio,
F.~A. 2003, \apj, 593, 772

\reference Freire, P.~C., Camilo, F., Lorimer, D.~R., Lyne, A.~G.,
Manchester, R.~N., \& D'Amico, N. 2001a, \mnras, 326, 901

\reference Freire, P.~C., Kramer, M., Lyne, A.~G., Camilo, F., Manchester,
R.~N., \& D'Amico, N. 2001b, \apj, 557, L105

\reference Freire, P.~C., Camilo, F., Kramer, M., Lorimer, D.~R., Lyne,
A.~G., Manchester, R.~N., \& D'Amico, N. 2003, \mnras, 340, 1359

\reference Freire, P.~C., Gupta, Y., Ransom, S.~M., \& Ishwara-Chandra,
C.~H. 2004, \apj, 606, L53

\reference Freire, P.~C., Hessels, J.~W.~T., Nie, D.~J., Ransom, S.~M.,
Lorimer, D.~R., \& Stairs, I.~H. 2005, \apj, in press (astro-ph/0411160)

\reference Fruchter, A.~S., \& Goss, W.~M. 2000, \apj, 536, 865

\reference Gao, B., Goodman, J., Cohn, H., \& Murphy, B. 1991, \apj,
370, 567

\reference Grindlay, J.~E., Heinke, C.~O., Edmonds, P.~D., Murray, S.~S.,
\& Cool, A.~M. 2001, \apj, 563, L53

\reference Grindlay, J.~E., Camilo, F., Heinke, C.~O., Edmonds, P.~D.,
Cohn, H., \& Lugger, P. 2002, \apj, 581, 470

\reference Guhathakurta, P., Webster, Z.~T., Yanny, B., Schneider, D.~P.,
\& Bahcall, J.~N. 1998, AJ, 116, 1757

\reference Harris, W.~E. 1996, AJ, 112, 1487; February 2003 revision
\\ ({\tt http://physwww.mcmaster.ca/\%7Eharris/mwgc.dat})

\reference Heggie, D.~C. 1975, MNRAS, 173, 729

\reference Heggie, D.~C., \& Rasio, F.~A. 1996, MNRAS, 282, 1064

\reference Heggie, D.~C., Hut, P., \& McMillan, S.~L.~W. 1996, \apj,
467, 359

\reference Heinke, C.~O., Grindlay, J.~E., Edmonds, P.~D., Lloyd, D.~A.,
Murray, S.~S., Cohn, H.~N., \& Lugger, P.~M. 2003, ApJ, 598, 516

\reference Heinke, C.~O., Grindlay, J.~E., Edmonds, P.~D., Cohn, H.~N.,
Lugger, P.~M., Camilo, F., Bogdanov, S., \& Freire, P.~C. 2005, \apj,
in press

\reference Hessels, J.~T.~W., Ransom, S.~M., Stairs, I.~H., Kaspi, V.~M.,
Freire, P.~C.~C., Backer, D.~C., \& Lorimer, D.~R. 2004, in IAU Symp.\
218: Young Neutron Stars and Their Environments, eds.\ F. Camilo \&
B.~M. Gaensler (San Francisco: ASP), p.\ 131

\reference Hut, P., McMillan, S., Goodman, J., Mateo, M., Phinney,
E.~S., Pryor, C., Richer, H.~B., Verbunt, F., \& Weinberg, M. 1992,
PASP, 104, 981

\reference Ivanova, N., Rasio, F.~A., Lombardi, J.~C., Jr., Dooley,
K.~L., \& Proulx, Z.~F. 2005, ApJ, in press

\reference Jacoby, B.~A., Chandler, A.~M., Backer, D.~C., Anderson,
S.~B., \& Kulkarni, S.~R. 2002, IAUC 7783

\reference Johnston, H.~M., Kulkarni, S.~R., \& Phinney, E.~S. 1992,
in X-Ray Binaries and Recycled Pulsars, eds.\ E.~P.~J. van den Heuvel \&
S.~A. Rappaport (Dordrecht: Kluwer), p.\ 349

\reference Jord{\' a}n, A., C{\^ o}t{\' e}, P., Ferrarese, L., Blakeslee,
J.~P., Mei, S., Merritt, D., Milosavljevi{\' c}, M., Peng, E.~W., Tonry,
J.~L., \& West, M.~J. 2004, \apj, 613, 279

\reference King, A.~R., Davies, M.~B., \& Beer, M.~E. 2003, MNRAS, 345, 678

\reference Krolik, J.~H., Meiksin, A., \& Joss, P.~C. 1984, \apj, 282, 466

\reference Kulkarni, S.~R., \& Anderson, S.~B. 1996, in IAU Symp. 174:
Dynamical Evolution of Star Clusters: Confrontation of Theory and
Observations, eds.\ P. Hut \& J. Makino (Dordrecht: Kluwer), p.\ 181

\reference Kulkarni, S.~R., Anderson, S.~B., Prince, T.~A., \& Wolszczan,
A. 1991, Nature, 349, 47

\reference Kumar, P., \& Goodman, J. 1996, \apj, 466, 946

\reference Leonard, P.~J.~T. 1989, AJ, 98, 217

\reference Lorimer, D.~R., Camilo, F., Freire, P., Kramer, M., Lyne,
A.~G., Manchester, R.~N., \& D'Amico, N. 2003, in ASP Conf.\ Ser.\ Vol.\
302: Radio Pulsars, eds.\ M. Bailes, D.~J. Nice \& S.~E. Thorsett (San
Francisco: ASP), p.\ 363

\reference Lyne, A.~G., Brinklow, A., Middleditch, J., Kulkarni, S.~R.,
\& Backer, D.~C. 1987, Nature, 328, 399

\reference Lyne, A.~G., Biggs, J.~D., Harrison, P.~A., \& Bailes, M. 1993,
Nature, 361, 47

\reference Lyne, A.~G., Manchester, R.~N., \& D'Amico, N. 1996, \apj,
460, L41

\reference Lyne, A.~G., Mankelow, S.~H., Bell, J.~F., \& Manchester,
R.~N. 2000, \mnras, 316, 491

\reference McConnell, D., Deshpande, A.~A., Connors, T., \& Ables,
J.~G. 2004, \mnras, 348, 1409

\reference McMillan, S., \& Hut, P. 1994, \apj, 427, 793

\reference McMillan, S.~L.~W., Taam, R.~E., \& McDermott, P.~N. 1990,
\apj, 354, 190

\reference Orosz, J.~A., \& van Kerkwijk, M.~H. 2003, \aap, 397, 237

\reference Phinney, E.~S. 1992, Phil.\ Trans.\ Roy.\ Soc.\ A, 341, 39

\reference Phinney, E.~S. 1993, in ASP Conf.\ Ser.\ Vol.\ 50: Structure
and Dynamics of Globular Clusters, eds.\ S.~G.\ Djorgovski \& G.\ Meylan
(San Francisco: ASP), p.\ 141

\reference Phinney, E.~S. 1996, in ASP Conf.\ Ser.\ Vol.\ 90: The
Origins, Evolutions, and Destinies of Binary Stars in Clusters, eds.\
E.~F.\ Milone \& J.-C.\ Mermilliod (San Francisco: ASP), p.\ 163

\reference Phinney, E.~S., \& Sigurdsson, S. 1991, Nature, 349, 220

\reference Podsiadlowski, P., Rappaport, S., \& Pfahl, E.~D. 2002, \apj,
565, 1107

\reference Pooley, D., et al. 2003, \apj, 591, L131

\reference Possenti, A., D'Amico, N., Manchester, R.~N., Sarkissian,
J., Lyne, A.~G., \& Camilo, F. 2001, astro-ph/0108343

\reference Possenti, A., D'Amico, N., Manchester, R., Camilo, F.,
Lyne, A.~G., Sarkissian, J., \& Corongiu, A. 2003, \apj, 599, 475

\reference Prince, T.~A., Anderson, S.~B., Kulkarni, S.~R., \& Wolszczan,
A. 1991, \apj, 374, L41

\reference Pylyser, E., \& Savonije, G.~J. 1988, A\&A, 191, 57

\reference Ransom, S.~M., Greenhill, L.~J., Herrnstein, J.~R., Manchester,
R.~N., Camilo, F., Eikenberry, S.~S., \& Lyne, A.~G. 2001, \apj, 546, L25

\reference Ransom, S.~M., Stairs, I.~H., Backer, D.~C., Greenhill, L.~J.,
Bassa, C.~G., Hessels, J.~W.~T., \& Kaspi, V.~M. 2004, \apj, 604, 328

\reference Ransom, S.~M., Hessels, J.~W.~T., Stairs, I.~H., Freire,
P.~C., Camilo, F., Kaspi, V.~M., \& Kaplan, D.~L. 2005, Science, in press

\reference Rappaport, S., Podsiadlowski, P., Joss, P.~C., Di Stefano,
R., \& Han, Z. 1995, MNRAS, 273, 731

\reference Rasio, F.~A. 2000, in IAU Colloq.\ 177, ASP Conf.\ Ser.\
Vol.\ 202: Pulsar Astronomy -- 2000 and Beyond, eds.\ M.\ Kramer, N.\
Wex \& R.\ Wielebinski (San Francisco: ASP), p.\ 589

\reference Rasio, F.~A. 2003, in ASP Conf.\ Ser.\ Vol.\ 302: Radio
Pulsars, eds.  M. Bailes, D.~J. Nice \& S.~E. Thorsett (San Francisco:
ASP), p.\ 385

\reference Rasio, F.~A., \& Heggie, D.~C. 1995, ApJ, 445, L133

\reference Rasio, F.~A., \& Shapiro, S.~L. 1991, \apj, 377, 559

\reference Rasio, F.~A., Pfahl, E.~D., \& Rappaport, S.~A. 2000, \apj,
532, L47

\reference Sabbi, E., Gratton, R.~G., Bragaglia, A., Ferraro, F.~R.,
Possenti, A., Camilo, F., \& D'Amico, N. 2003, A\&A, 412, 829

\reference Sepinsky, J.~F., Saffer, R.~A., Shara, M.~M., \& Zurek,
D. 2002, BAAS, Vol.\ 34, p.\ 1103

\reference Shara, M.~M., Bergeron, L.~E., Gilliland, R.~L., Saha, A.,
\& Petro, L. 1996, \apj, 471, 804

\reference Shara, M.~M., Saffer, R.~A., \& Livio, M. 1997, ApJ, 489, L59

\reference Sigurdsson, S. 2003, in ASP Conf.\ Ser.\ Vol.\ 302: Radio
Pulsars, eds.\ M. Bailes, D.~J. Nice \& S.~E. Thorsett (San Francisco:
ASP), p.\ 391

\reference Sigurdsson, S., \& Phinney, E.~S. 1993, \apj, 415, 631

\reference Sigurdsson, S., Richer, H.~B., Hansen, B.~M., Stairs, I.~H.,
\& Thorsett, S.~E. 2003, Science, 301, 193

\reference Spergel, D.~N. 1991, Nature, 352, 221

\reference Stappers, B.~W., Bailes, M., Manchester, R.~N., Sandhu, J.~S.,
\& Toscano, M. 1998, ApJ, 499, L183

\reference Taam, R.~E., \& Sandquist, E.~L. 2000, ARAA, 38, 113

\reference Thorsett, S.~E., Arzoumanian, Z., McKinnon, M.~M., \& Taylor,
J.~H. 1993, \apj, 405, L29

\reference Thorsett, S.~E., Arzoumanian, Z., Camilo, F., \& Lyne,
A.~G. 1999, \apj, 523, 763

\reference van Kerkwijk, M.~H., Kaspi, V.~M., Klemola, A.~R., Kulkarni,
S.~R., Lyne, A.~G., \& Van Buren, D. 2000, \apj, 529, 428

\reference van der Sluys, M.~V., Verbunt, F., \& Pols, O.~R. 2004, A\&A,
in press (astro-ph/0411189)

\reference Verbunt, F. 1987, \apj, 312, L23

\end{references}
\end{document}